\documentclass[10pt,journal,compsoc]{IEEEtran}
\ifCLASSOPTIONcompsoc
  \usepackage[nocompress]{cite}
\else
  \usepackage{cite}
\fi
\usepackage{comment}
\usepackage{blindtext}
\usepackage{graphicx}
\usepackage{subfig}
\usepackage{caption}
\usepackage{comment}
\usepackage{booktabs,amsfonts,dcolumn}
\usepackage{algorithm}
\usepackage{algpseudocode}
\usepackage[dvipsnames]{xcolor}
\usepackage{amsmath}
\usepackage{array}

\ifCLASSINFOpdf
\else
\fi
\begin{document}
%
\title{Scalable Hierarchical Instruction Cache for Ultra-Low-Power Processors Clusters}
%
%
%

\author{Jie Chen,
        Igor Loi,
        Eric Flamand,
        Giuseppe Tagliavini,~\IEEEmembership{Member,~IEEE,}
        Luca Benini,~\IEEEmembership{Fellow,~IEEE,}
        Davide~Rossi,~\IEEEmembership{Member,~IEEE,}
\IEEEcompsocitemizethanks{\IEEEcompsocthanksitem J. Chen, G. Tagliavini, L. Benini and D. Rossi are with University of Bologna, Italy. L. Benini is with ETH Zurich, Switzerland. J. Chen, I. Loi and Eric Flamand are with GreenWaves Technologies, Grenoble, France.\protect\\
E-mail: jie.chen2@studio.unibo.it }
\thanks{This work was supported in part by EU Horizon 2020 Research and Innovation projects The European Pilot under Grant 101034126, in part by WiPLASH under Grant 863337, in part by ECSEL Horizon 2020 project AI4DI under Grant 826060 and in part by GreenWaves Technologies.}}

\IEEEtitleabstractindextext{%
\begin{abstract}
High Performance and Energy Efficiency are critical requirements for Internet of Things (IoT) end-nodes. Exploiting tightly-coupled clusters of programmable processors (CMPs) has recently emerged as a suitable solution to address this challenge. One of the main bottlenecks limiting the performance and energy efficiency of these systems is the instruction cache architecture due to its criticality in terms of timing (i.e., maximum operating frequency), bandwidth, and power. We propose a hierarchical instruction cache tailored to ultra-low-power tightly-coupled processor clusters where a relatively large cache (L1.5) is shared by L1 private caches through a two-cycle latency interconnect. To address the performance loss caused by the L1 capacity misses, we introduce a next-line prefetcher with cache probe filtering (CPF) from L1 to L1.5. We optimize the core instruction fetch (IF) stage by removing the critical core-to-L1 combinational path. We present a detailed comparison of instruction cache architectures' performance and energy efficiency for parallel ultra-low-power (ULP) clusters. Focusing on the implementation, our two-level instruction cache provides better scalability than existing shared caches, delivering up to 20\% higher operating frequency. On average, the proposed two-level cache improves maximum performance by up to 17\% compared to the state-of-the-art while delivering similar energy efficiency for most relevant applications. 
\end{abstract}

}

\maketitle

\IEEEdisplaynontitleabstractindextext

%
\IEEEpeerreviewmaketitle

\IEEEraisesectionheading{\section{Introduction}\label{sec:introduction}}

\IEEEPARstart{W}{ith} the increasing demand for edge computing capabilities driven by the momentum of the Internet of Things (IoT), performance and power consumption are becoming stringent constraints for edge computing platforms. On the one hand, increasing performance reduces the processing latency for intensive workloads with strict timing constraints (e.g., neural network inference for drone navigation \cite{8264734}). On the other hand, low-power operation extends the battery lifetime, guaranteeing low maintenance costs and lower environmental impact.

In recent years, multi-core architectures have been proposed to improve the computational capabilities of edge systems using parallel processing architectures, limiting the operating frequency to that available in the near-threshold operating region to combine performance with energy efficiency \cite{b6}. These architectures usually adopt a symmetric multiprocessing (SMP) approach characterized by multiple cores cooperating in the same memory space. SMP systems are common in the literature on computer design, also widely adopted in the embedded systems domain \cite{karam2009trends}. From a computational perspective, this template promotes the adoption of a programming approach where all cores execute the same code on different parts of the input dataset. This paradigm is referred to as single-program multiple-data (SPMD) \cite{flynn1972some}. In addition, these architectures can feature specialized hardware extensions to reduce the parallelization overhead (e.g., hardware-assisted synchronization), the memory footprint and the number of cycles per processed data item (e.g., packed-SIMD operations).

The instruction cache design plays a crucial role in obtaining high performance and energy efficiency in the parallel execution scenario. In high-performance multi-core clusters, instruction caches' area and power consumption are small, often negligible, compared to the large and power-hungry processors and fully-coherent data caches. In contrast, in the ultra-low-power domain, the contributions from data memories and cores are significantly reduced due to lower frequencies and smaller datasets. In this context, instruction cache increase is an important parameter and can consume up to 50\% of the total power and area, becoming one of the main efficiency bottlenecks \cite{b20}. In addition, their power consumption is largely influenced by memory implementation. P. Meinerzhagen et al. propose the adoption of latch-based standard cell memories (SCMs) \cite{b15} and a \textit{private model} (i.e., one instruction cache instance per core). This design does not introduce critical paths in the cache subsystem and guarantees a single-cycle access latency coupled with low power consumption. However, private caching implies that even if a single instruction stream is executed in SPMD, it will be cached multiple times (i.e., one cached copy for each core in the cluster). Thus, small configurations suffer from heavy miss penalties (up to 80\% performance degradation with 256B instruction caches compared to an ideal cache) \cite{b7}. At the other extreme, for configurations with large capacity, the multiple private I-caches tend to dominate the cluster area \cite{b7}.

To tackle the issues described above, Loi et al. \cite{b7} proposed two alternative designs of \textit{shared} multi-banked instruction caches that avoid multiple private copies, thereby offering a larger capacity when running SPMD code (i.e., $8\times$ w.r.t. a private model, where $8$ is the number of cores in the target platform). The single-port (\textit{SP}) shared cache introduces connections between each core and all the cache banks. This solution still guarantees a single-cycle access latency but suffers from congestion when several cores simultaneously access the same cache bank. Unfortunately, this adverse condition is not sporadic but frequently occurs in executing parallel applications based on the SPMD paradigm since cores fetch the same code by construction. Moreover, this design introduces a critical path between the cores and the cache banks through the interconnect, impacting the maximum achievable frequency and limiting performance accordingly. Multiple-Ported (MP) shared caches are a brute-force alternative to reduce the number of cycles required to access the banks, relieving the congestion issue. Nevertheless, this solution severely impacts the cache area, so it is suitable only for small cache capacities. Last but not least, also MP caches suffer from timing issues when the number of cores scales up.

To solve the scalability issue of the shared caches, we introduce a hierarchical instruction cache design targeting the scalability challenges of ultra-low-power multi-core architectures. The proposed instruction cache is still based on latch-based memories to improve energy efficiency and voltage scalability over power-hungry SRAM memories \cite{b15}. This hierarchical approach, along with optimization to the cluster's cores and adoption of well-known prefetching techniques, reduces both access conflicts and critical path while guaranteeing a low miss rate. To guarantee a minimal impact on the area, the capacity of this cache level is limited by design (e.g., to 512 B). Introducing an additional cache level increases the total capacity to that of the other shared cache solutions (e.g., 4 KB).

The main contributions of our work are:

\begin{itemize}
  \item The design of a two-level instruction cache combining private (L1) caches with a shared (L1.5) cache to relax the timing constraints while achieving an expected latency very close to single-cycle. The proposed architecture marries the benefits of traditional private caches (fast design with a short critical path) with the capacity benefit of shared caches for SPMD programs.
  \item A 128-bit next-line prefetching unit with cache probe filtering (CPF) in the L1 and out-of-order L1-to-L1.5 interconnect to recover the performance drop caused by the L1 capacity miss. The prefetching unit can be enabled or disabled by software to fit different application requirements.
  \item An optimization of the core instruction fetch stage (IF) to better fit the functional characteristics of the proposed hierarchical cache and to reduce the latency from the cores to L1 caches, improving the operating frequency by 15\% w.r.t. the baseline implementation.
  \item Synthesis and implementation of the proposed solution in GF 22nm FD-SOI technology. The cache modules are integrated into the Parallel Ultra-Low-Power Platform (PULP), an 8-core cluster based on an SMP architectural template. We performed an extensive and comprehensive exploration using a set of synthetic benchmarks to compare our solution with state-of-the-art alternative designs in terms of area, power, performance, and energy efficiency.
 \item The experimental assessment of a set of real-life applications in the areas of digital signal processing (DSP) and convolutional neural networks (CNN) to evaluate the performance and energy efficiency.
\end{itemize}

The proposed architecture overcomes the main bottlenecks of existing shared caches solutions, providing 20\% higher operating frequency when integrated on a tightly-coupled cluster of RISC-V cores, and better scalability. Indeed, when moving from a baseline 8-core cluster to a 16-cores cluster, the operating frequency decreases by only 7\%, from 22\% to 26\% higher than 16-cores clusters built around shared caches. On average, when executing a set of real-life low-miss rate IoT applications on an 8-core cluster, the proposed two-level cache improves maximum performance by up to 17\% compared to the state-of-the-art, delivering similar energy efficiency.

\section{Related Work}

\subsection{Instruction Memory in ULP SoCs}

The instruction fetch hierarchy is one of the most critical issues for ultra-low-power systems-on-chip as it can consume up to 50\% of the overall system energy\cite{b20}. A standard methodology to reduce energy consumption is to adopt advanced memory technologies. A few notable examples are presented below.

Oboril et al. \cite{b19} show that a hybrid combination (SRAM for the L1-data cache, SOT-MRAM for both L1 instruction cache and L2 cache) can reduce the energy consumption by 60\% while the performance increases by 1\% compared to an SRAM-only configuration, targeting a 65 nm technology node. Kuan and Adegbija \cite{8721116} show that an energy-efficient, highly adaptable last-level STT-RAM cache can reduce the average energy consumption by 60\% in a quad-core system while introducing marginal latency overhead. Myers et al. \cite{b21} introduce a Cortex M0+ based system with two 4KB 10T SRAM optimized for sub-threshold operation. Ickes and al. \cite{b22} present a 10 pJ/cycle 32-bit microprocessor SoC with 1 KB (8 x 128B) instruction cache based on latch-based SCMs \cite{b15} \cite{b16}. 

Replacing or combining state-of-the-art memory technologies is one of the most efficient ways to reduce instruction cache power consumption and improve system energy efficiency. PULPv2 \cite{b6} uses SCMs instead of SRAMs in the instruction cache, increasing the SoC energy efficiency by 38\%. SCMs present extremely interesting features for small memory size, low-voltage, and energy-efficient designs, since (i) they can operate with very low voltage, even lower than 10T SRAMs optimized for low voltage \cite{b15}, and (ii) their energy per access is significantly smaller than SRAMs. Nevertheless, although the controlled placement of standard cells memory array reduces area overhead \cite{b16}, there is still 2$\times$ area penalty with respect to the same size SRAMs-based memory. Thus, it is clear that since there is a solid motivation to use energy-efficient but low-density memories for instruction caching, there is a strong push to maximize the capacity of the caches through sharing schemes.

\subsection{Improving Instruction Fetch Efficiency}

The idea of sharing memory is not new, and the trade-off between shared and private caches is well-known \cite{origin}. For instance, in the context of data caches for high-end cache coherent systems, Chun Liu et al. \cite{Liu_2004} proposed a shared L2 cache architecture improving performance by more than 40\% over the private organization. While this computational pattern is common in general-purpose architectures, a more common computational pattern in digital signal processing is data-parallelism with multi-banked scratchpad memories. A notable example of high-end computing platforms exploiting this pattern is that of General Purpose Graphic Processing Units (GP-GPU). From an instruction fetch perspective, in GP-GPUs, the compute units in each multiprocessor execute their threads in lock-step following the order issued by the instruction dispatcher, which is shared among all of them \cite{gpu}. However, Single Instructions Multiple Threads (SIMT) architectures like GP-GPUs are much less flexible than SPMD architectures. Executing control code and data-dependent conditional code causes significant performance degradation and efficiency since thread divergence is avoided by sequentializing the execution of the conflicting parts. For instance, a code executing an \emph{if-then-else} statement will force the dispatcher to execute the \emph{then} part first, and then proceed to the \emph{else} part. Unfortunately, this is a common scenario in many IoT applications. 

Dynamic cache reconfiguration (DCR) is another effective technique to optimize energy consumption in many-core architecture \cite{Zhang_2005}\cite{Wang_2011}. Dynamically tuning the appropriate cache parameters (such as associativity and cache line size) can satisfy the memory access behavior of different applications to improve the cache fetch efficiency and save a significant amount of energy. In our work, we propose an orthogonal approach, fixing cache line and associativity parameters at design time for low-area overhead while exploiting instruction cache sharing to leverage SPMD nature of the target applications. The fully balanced shared-memory solutions proposed by Loi et al. \cite{b7} leverage the larger capacity of shared caches exploiting the SPMD pattern of the applications. However, in \cite{b7}, as shared cache banks are connected to the cores through a single-cycle interconnect or multiple ports, long paths and heavily congested design limit the frequency when the number of cores and memories scale up. In this work, we overcome these limitations by proposing a hierarchical, low-latency design that improves the operating frequency and system performance both for typical and large cluster configurations (i.e., 8- and 16-cores).

\subsection{Instruction Cache Prefetching}

Prefetching is a conventional approach to mitigate the impact of cache misses. Next-line prefetching \cite{origin} is an efficient method to improve performance for sequential execution, loading the block of instructions following the current one in memory. However, it is unsuitable for non-sequential execution paths caused by jumps, conditional branches, and system calls. Despite these shortcomings, it is still an effective strategy to reduce cache misses by 20-50\%. 

Fetch-directed instruction prefetching (FDP) \cite{Reinman_1999} separates the branch predictor and the instruction cache so that the branch predictor can run ahead of the instruction cache fetch. Using Cache Probe Filtering (CPF) to remove useless prefetch requests in the Fetch Target Queue (FTQ), the branch fetch blocks can be predicted accurately, thereby saving bus bandwidth to the L2 cache. FDP relies on accurate branch predictors and a sufficiently large Branch Target Buffer (BTB) to cover the control flow.

Temporal prefetching \cite{Ferdman_2008}-\cite{Ferdman_2011} is based on the fact that the stream of instruction cache misses is repetitive and eliminates the future instruction cache misses directly by tracing these temporally correlated streams. Based on that, RAS-directed instruction prefetching (RDIP) \cite{Kolli_2013} correlates instruction cache misses with the program context captured from the Return Address Stack (RAS). It stores these misses in a Miss Table that is looked up using the signatures formed from the contents of the RAS. However, the main shortcoming of temporal prefetching is its high storage budget requirements, larger than 60 KB. From an industrial perspective, Ishii et al. \cite{Ishii_2021} propose an effective FDP-base frontend design with only 195 bytes of hardware overhead. It has two enhancements, taken-only branch target history and post-fetch correction, to overcome its previous issues. This solution outperforms the 1st Instruction Prefetching Championship (IPC-1) winners with a 128 KB storage budget.

\subsection{Discussion}

In general, most state-of-the-art two-level caches and prefetching techniques are designed for high-performance processors operating at high frequency (i.e., $>$ 1 GHz), where large SRAM-based L1 caches work in conjunction with complex branch predictors to hide the large refill latency towards L2 memories. While the SRAM-based caches and complex prefetching techniques require negligible area overheads when integrated into high-performance processors due to the large silicon area of other blocks, they might become area- and power-dominant in a ULP context if not adequately managed at the system level. In the proposed work, we employ a 4~kB energy-efficient \cite{b16} latch-based shared cache coupled with small 512~B L1 caches, also implemented with latches. L1 caches are coupled to a simple ($<$ 1~kGates) sequential next-line (128-bit) prefetcher to hide the 2-cycle access latency to the L1.5 memory (improving performance by 7\% on average) without jeopardizing the area and energy efficiency of the system. Furthermore, we integrate cache probe filtering (CPF) to improve the efficiency of L1 to L1.5 bandwidth and an out-of-order interconnect to maximize L1 to L1.5 bandwidth.

Finally, it is noteworthy to mention that most recent work on instruction caches for multi-core architectures proposes alternative replacement policies suitable for specific application domains \cite{diaz2021near} \cite{ghosh2021srcp} or interference-free techniques for real-time systems \cite{xiao2022cache} \cite{cabo2021safesu}. Our work is focused on the architectural design and practical trade-offs of integrating a hierarchical cache in an ultra-low-power multi-core platform.

\section{Background} \label{sec:background}

\graphicspath{ {figures/} }
\subsection{Cluster Architecture} \label{sec:soc}

The baseline system we consider in this work is a tightly coupled cluster of processors dedicated to parallel acceleration, including eight 32-bit RISC-V cores (RI5CY \cite{b11}). RI5CY was selected as baseline core for the proposed architecture thanks to its high energy efficiency coupled and its open source nature allowing to easily modify the internal micro-architecture, contrarily to closed source cores such as ARM Cortex M4. The RISC-V core is based on an in-order, single-issue, four-stage pipeline micro-architecture without branch prediction, improved with extensions targeting energy-efficient near-sensor data analytics. These extensions include hardware loops, load/store instructions with pre/post increment, SIMD operations, with the aim to increase throughput and energy efficiency in parallel signal processing workloads \cite{b11}. No data cache is present; the architecture uses explicit Direct memory access (DMA) transfers to move data between L1 and L2 memory avoiding memory coherency overhead and additional area and power penalties \cite{b14}. Thus, the data memory is implemented as a word-level interleaved multi-bank Tightly Coupled Data Memory (TCDM) \cite{Rahimi_2011}. The cores share a 128~KB single-cycle latency TCDM and a DMA coupled to the TCDM as well \cite{b4}. A banking factor of 2 is kept to reduce banking conflict probability. The instruction cache and DMA are connected to an AXI4 cluster bus for fetching off-cluster data. To increase the throughput of each core, a next-line prefetch buffer of one cache line (16~B for 128-bit interface) is instantiated between each core and the instruction cache. Finally, thanks to the low-voltage SRAMs for L2 memory (1~MB) of the SoC domain and cluster TCDM, the supply voltage of the SoC and cluster can scale down to near-threshold 0.65~V in the 22~nm FD-SOI target technology. In the SoC domain, a Fabric Controller (FC) core is responsible for executing the sequential tasks and orchestrating the cluster parallel acceleration.

\begin{figure}[t!]
\includegraphics[width=0.48\textwidth]{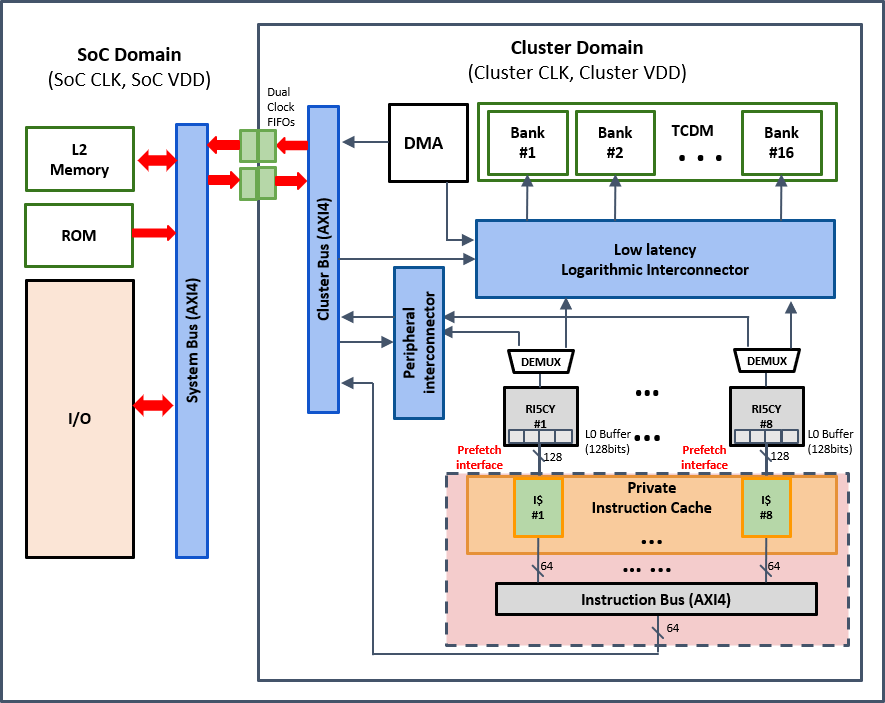}
\caption{Baseline SoC architecture, featuring an 8-core cluster with private instruction cache, the cache is shown in green.}
\label{fig:soc}
\end{figure}

\subsection{Legacy I\$ Architectures}

This section introduces the different instruction cache architectures considered as a baseline for the exploration performed in the paper. We thereby describe more traditional private caches as well as two different types of shared caches. All the variants described in the following are implemented with SCMs for improved energy efficiency in near-threshold processors. Although the analysis was performed using RI5CY core, similar results would be expected with other low-power processors with shallow pipelines such as IBEX or ARM Cortex M4.

\subsubsection{Private cache}

\begin{figure}[t!]
\begin{center}
\includegraphics[width=0.3\textwidth]{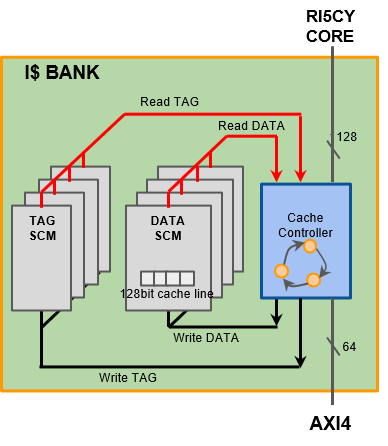}
\end{center}
\vspace{-0.5cm}
\caption{Cache bank subsystem}
\label{fig:cache_bank}
\end{figure}

The baseline cluster features private instruction caches (Fig.~\ref{fig:cache_bank}). Each private cache bank comprises three elements: the TAG array, the DATA array, and a cache controller. The core's 128-bit next-line prefetcher described in the previous section exploits a request-grant handshake protocol to fetch 128-bit cache lines through the controller. A pseudo-random (PRAND) policy is used for replacement. If a cache miss occurs, the cache fetches a new cache line from L2 memory through the 64-bit AXI4 bus. Private caches are typically fast (i.e., one-cycle latency) and simple (i.e., low-power) \cite{Liu_2004}. On the other hand, data replication and high miss penalty are the major drawbacks of the private instruction cache. This aspect, along with the latch-based implementation that improves energy efficiency but increases the area \cite{b16}, leads to a decrease in their performance, area efficiency, and energy efficiency when dealing with SPMD workloads in tightly coupled clusters as soon as the footprint of the application exceed the size of the L1 cache \cite{b1}.

\begin{figure*}[h!]
\centering
\subfloat[Single-port shared cache\label{fig:SP}]{
  \includegraphics[width=0.4\textwidth]{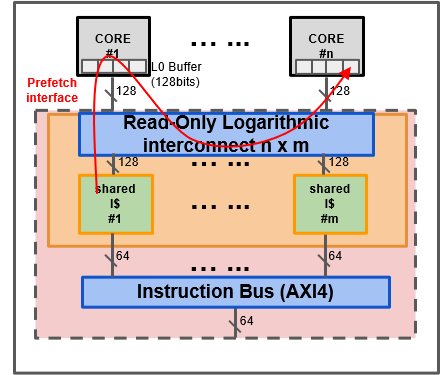}
}
\hspace{5em}
\subfloat[Multi-port shared cache\label{fig:MP}]{
  \includegraphics[width=0.38\textwidth]{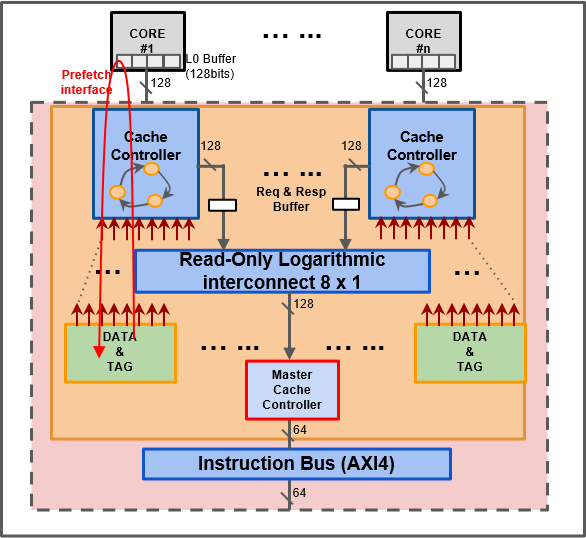}
}
\caption{Shared instruction cache architectures and their critical paths shown in the red arrow.}
\label{fig:shared_caches}
\end{figure*}

\subsubsection{Single-Port Shared Cache}

To overcome the described limitations of private caches, two different shared cache architectures have been presented by Loi et al. \cite{b1}. The first one, called Single-Port (SP) shared cache, is shown in Fig.~\ref{fig:SP}. It leverages a multi-banked cache architecture connected to the instruction fetch stage of the cores through a low-latency logarithmic interconnect similar to the one described in \cite{Rahimi_2011}. SPMD applications can benefit from the fact that the whole cache is shared among all the cores. Hence, all the cores see on their address space the whole cache capacity as opposed to private caches, where each core can only use the whole capacity divided by the number of cores. On the other hand, the interconnect between the cores and the cache banks poses considerable timing pressure on the cluster being on its critical path. More precisely, the critical path starts from the output of the cache bank's tags back to the IF stage of the core through the response tree of the interconnect and then again in the arbitration tree of the interconnect towards the IF stage of the other cores. This long path is necessary for guaranteeing the single-cycle behavior of the network. Since each cache bank can serve one refill request at a time, this design also causes congestion when several cores access the same cache banks on the same cycle for parallel applications \cite{b1}.

\subsubsection{Multiple-port Shared Cache}

To overcome some of the limitations of SP cache, reducing congestion to the cache banks, a second architecture was proposed in \cite{b1}. The Multi-Port cache exploits a few memory banks, still based on latches, featuring multiple private read ports, one per core. TAG and DATA memories are shared while keeping the cache controllers private for each core and close to the core fetch interfaces (Fig.~\ref{fig:MP}). This architecture solves the access congestion issue of SP, since each core has its own read port on the shared banks. Moreover, the critical path of the design does not cross a large interconnect as in the case of SP. On the other hand, TAG and DATA memories with a large number of ports cause significant placement and routing congestion during physical implementation, leading to severe timing, area, and power overheads. As a result, it is suitable only for cache sizes up to a few KB.



\section{Two-Level instruction cache}

\begin{figure}[t!]
\begin{center}
\includegraphics[width=0.35\textwidth]{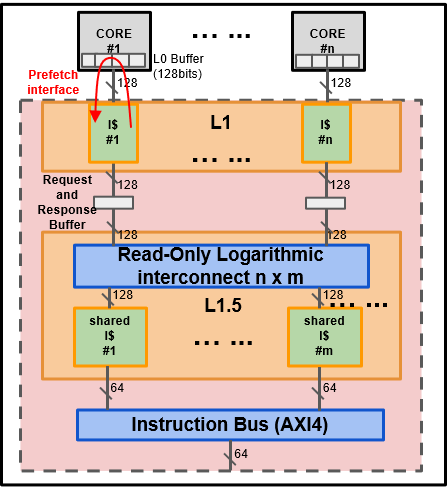}
\end{center}
\caption{Two-level Instruction cache combines private cache with single read-port shared cache. The critical paths shown in the red arrow.}
\label{fig:HIER}
\vspace{-0.5cm}
\end{figure}

The goal of the proposed work is to create an instruction cache architecture overcoming the physical implementation issues (long critical path and high routing congestion) of SP and MP architectures given by the presence of the long interconnect for SP (Fig.~\ref{fig:SP}) and multiple read ports for MP (Fig.~\ref{fig:MP}), respectively, still being able to tolerate the pressure of large-footprint applications. To accommodate these requirements, we propose a two-level instruction cache, described in Fig.~\ref{fig:HIER} composed of a small private cache (L1) tightly coupled to a larger shared cache (L1.5), similarly to the SP, connected through a single clock latency interconnect. This approach cuts the long critical paths from the core to the interconnect and back to the core (Fig.~\ref{fig:SP}) while benefiting from the low latency access time from the core to the L1 and from L1 to L1.5.

\begin{figure*}[h!]
\centering
\subfloat[Two-level cache with L1 prefetch, the critical path is highlighted by the red arrow, arbiter and out of order interconnect is highlighted.\label{fig:HIER_PRE}]{
  \includegraphics[width=0.35\textwidth]{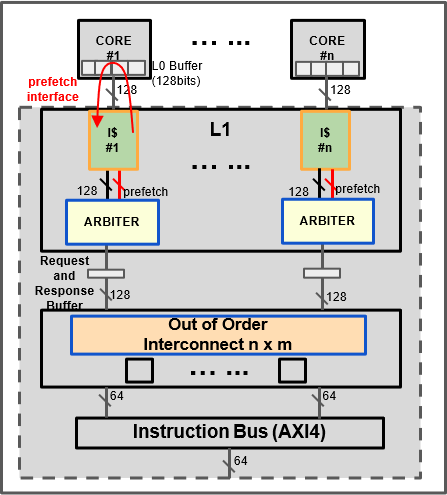}%
} \hspace{3em}%
\subfloat[L1 cache bank with an additional prefetch control unit and dual-port latched-based TAG memories for TAG Lookup.\label{fig:prefetch}]{
   \includegraphics[width=0.4\textwidth]{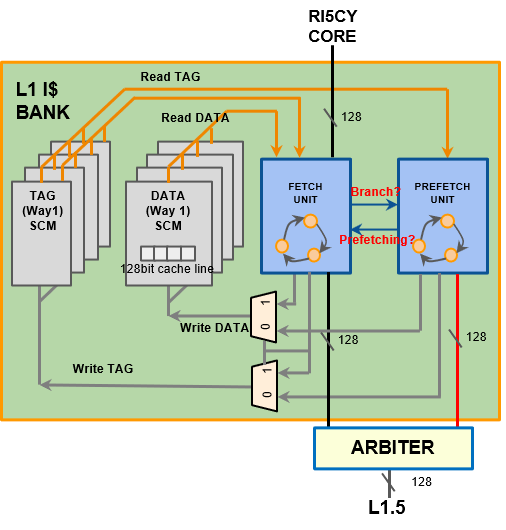}
}
\caption{Two-level cache with L1 next-line prefetching}
\label{fig:prefetch_all}
\vspace{-0.5cm}
\end{figure*}

As shown in Fig.~\ref{fig:HIER}, there are 8 private caches, and each features $1/8$ cache capacity of SP. To reduce the L1 cache miss rate, the L1.5 features the same cache capacity as SP. Unlike the 8 shared banks adopted by SP, the L1.5 cache has only 2 shared banks with an 8x2 logarithmic interconnect to reduce the area since the critical path is not from the L1 cache to the L1.5 cache. This approach significantly reduces the complexity of the interconnect removing the critical path present in the SP. Moreover, it does not suffer the congestion issues related to the multiple ports of the MP. The width of the fetch interface and the cache line are 128 bits. Between the L1 and L1.5, the proposed hierarchical cache features an optional request buffer and a response buffer (i.e., it can be enabled with a System Verilog parameter) to cut potentially critical paths from L1 to L1.5. Since the response buffer is sufficient in the presented cluster implementation to avoid critical paths from L1 to L1.5, the request buffer has been disabled in the experiments performed in this work. On the other hand, this buffer is a powerful knob to improve the system scalability towards high-end clusters optimized for frequency or featuring a larger number of cores, taking advantage of the hierarchical structure of the cache.

In the proposed architecture, the access time of L1 and L1.5 is one cycle and two cycles, respectively. In the case of banking conflicts in the L1.5 cache, the access time can be larger than two cycles depending on the number of parallel requests. However, the contention on the L1.5 banks decreases significantly with respect to \textit{SP} thanks to the presence of the L1 banks filtering many requests to L1.5. Compared to a private cache, the two-level cache performance can improve largely for high-footprint SPMD applications as it avoids replication in the L1.5 cache, thereby increasing actual capacity. Still, the presence of relatively small single-cycle L1 caches may lead to an increase in cycle count to execute a program with respect to the shared cache architectures \cite{DATE_Jie}. To overcome this issue, in this work, we propose a cache prefetching between the L1 and L1.5.

\begin{figure}[t!]
\begin{center}
\includegraphics[width=0.48\textwidth]{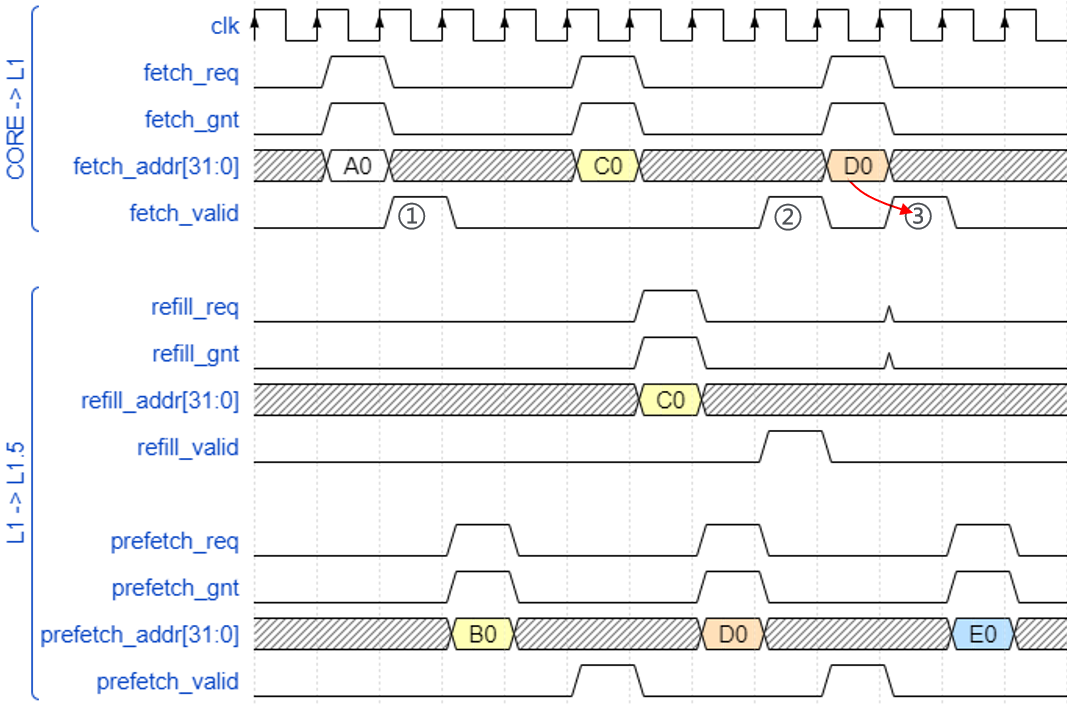}
\end{center}
\caption{Timing diagram of L1 to L1.5 prefetch. The upper timing diagram describes the core fetching from L1 with one cycle latency when hit (first fetch) and three cycles latency when miss and hit in the L1.5 (second fetch). The bottom timing diagram describes L1 refill and prefetch to the L1.5 with two cycles latency when hit. Once there is a core fetch, prefetch starts in the next cycle.}
\label{fig:prefetch_timing}
\end{figure}

\subsection{L1 to L1.5 Next-line Prefetching}

Instead of introducing much extra storage budget, we use a simple L1 next-line (128-bit cache line) prefetching (Fig. \ref{fig:prefetch_all}) to deal with L1 capacity misses. Since the prefetch-on-miss is insufficient to hide the latency for small L1, we always use prefetching. Once there is a fetch request from the core, a prefetch request is issued immediately. To avoid redundant prefetching, we use cache probe filtering (CPF) with the help of dual-port TAG memories, implemented with latches for parallel fetch and prefetch cache LOOKUP (Fig.~\ref{fig:prefetch}). Thanks to the CPF, the bandwidth from L1 to L1.5 slightly increases only when non-sequential sequences occur.

To avoid cache pollution, we store only valid prefetch cache lines in the cache. When performing the cache lookup, if the fetch address hits the prefetch buffer (shown by the signal \textit{Branch} in Fig.~\ref{fig:prefetch}), the prefetch buffer responds to the core directly. Besides, if the target prefetching is not finished, the fetch unit waits until the end of the prefetching (shown by the signal \textit{Prefetching} in Fig.~\ref{fig:prefetch}) and writes the prefetch cache line to the cache.

Nevertheless, there is a speed mismatch between core fetch and prefetch due to the different refill latency of each memory level. As a result, two different conditions might occur, shown in Fig.~\ref{fig:prefetch_timing}:

\begin{itemize}
\item Prefetch is faster than core fetch, shown in the first $fetch\_valid$. In this case, prefetch waits for the next valid core fetch to trigger again. If the next core fetch hits, we say prefetch succeeds. If not, we know that there is a branch. The prefetch control unit waits for the branch's valid fetch and restarts from the new address shown in the second $fetch\_req$. The branch address is 0xC0 instead of 0xB0, and prefetch restarts from next address - 0xD0.
\item Prefetch is slower than core fetch. Even though the prefetch is valid, core fetch still has a miss since prefetch is in doing. However, if the control logic knows that a valid prefetch is ongoing, the fetch can Wait for the Unfinished Prefetch (WUP) or directly use the prefetch data to save at least one cycle (red arrow) for TAG LOOKUP indicated by the signal $is\_prefetch$ in Fig.~\ref{fig:prefetch}. The third core fetch indicates a miss from the TAG LOOKUP while finding a valid prefetch, then the refill to L1.5 is canceled, and the cache responds directly.

\end{itemize}

\subsection{Out-of-order Interconnect} \label{lic_out_of_order}

In Fig.~\ref{fig:prefetch}, the fetch and prefetch control units issue requests to L1.5 cache through an arbiter to share the bandwidth, and this useful prefetch increase bandwidth a little only when branches happen. Instead of increasing 2$\times$ ports for the interconnect, which will bring more congestion and delay, we support out-of-order transfer for the fetch and prefetch. The prefetch can start directly without waiting for the unfinished fetch or vice versa. To identify the order of valid responses for fetch and prefetch from the shared banks, a transfer identifier field has been introduced in the arbiter. If the two responses come in the same cycle, we omit the prefetch data without interfering with the normal fetch. In real practice, when there is no miss in the L1 cache (always sequential fetch without branch), the bandwidth is almost 100\% occupied by the total 2$\times$8 requests from the fetch and prefetch. With the presented prefetch scheme, we improve the performance without influencing normal fetch with minimal area overhead, including the prefetch control unit, additional read port for TAG, and out-of-order interconnect to improve the performance and keep the energy efficiency. 

\begin{figure}[h]
\begin{center}
   \includegraphics[width=0.35\textwidth]{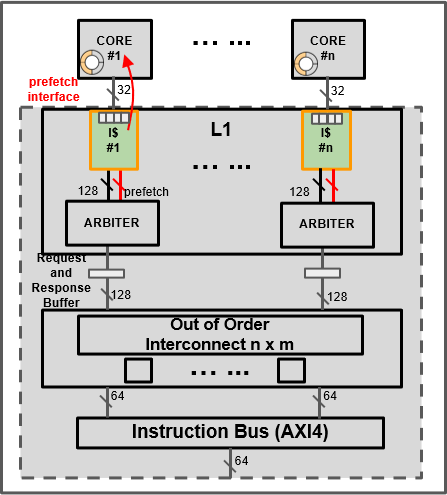}
\end{center}
\caption{Two-level cache with L1 prefetch after IF optimization with 4x32bit ring FIFO buffer and additional conditional branch pipeline. Critical path is shown by the red arrow.}
\label{fig:critical_opt}
\vspace{-0.5mm}
\end{figure}

\begin{figure}[h]
\begin{center}
\includegraphics[width=0.48\textwidth]{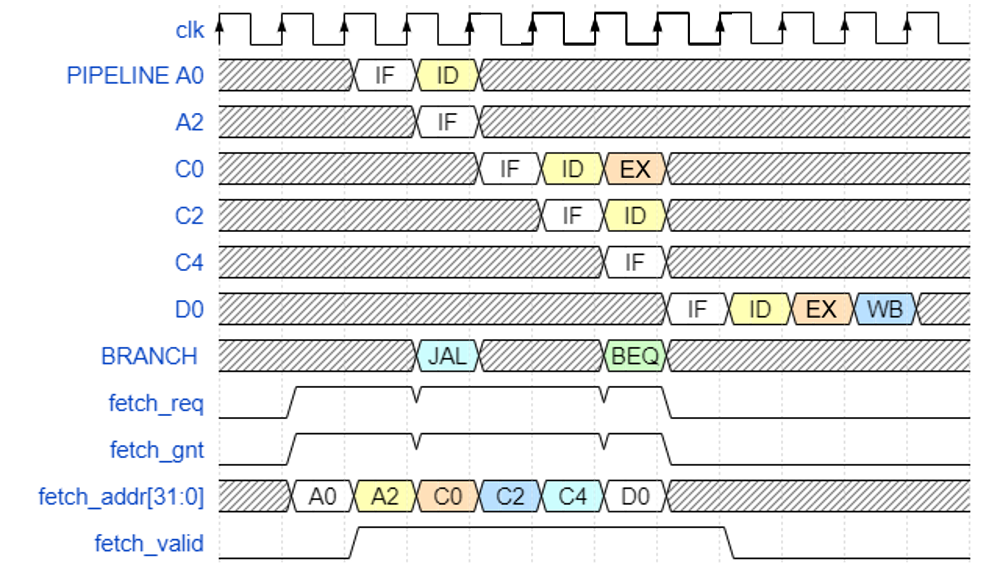}
\end{center}
\caption{Instruction fetch and branch in RI5CY core with 4-stage pipeline.}
\label{fig:branch}
\vspace{-0.5cm}
\end{figure}

\subsection{Instruction Fetch Stage Optimization} \label{if_opt}

As mentioned before, the caches with the legacy 128-bit instruction fetch stage have limited frequency due to some long combinational paths through the instruction fetch stage (IF) described below. There are two types of paths shown in Fig.~\ref{fig:HIER_PRE}: 1) The path from $fetch\_valid$ or $fetch\_data$ to $fetch\_req$ and back with $fetch\_gnt$ to the core. It exists because the IF does prefetch depending on the previous fetched (unaligned or non-unaligned, compressed or non-compressed) instruction to fetch the next instruction as soon as possible. Besides, the $fetch\_req$ is acknowledged by $fetch\_gnt$ back to the core. These port-to-port paths cause another critical path for the L1 cache, from the $fetch\_valid$ of the L1 cache's TAG or DATA arrays to $fetch\_req$, then back to it when TAG LOOKUP. This path gets worse when logarithmic interconnect is used, such as in SP cache. 2) The path from the instruction execution stage (EX) to the IF $fetch\_req$ until the cache has conditional branches. As shown in Fig.~\ref{fig:branch}, instruction 0xA0 is an unconditional jump taken directly from the ID stage to instruction 0xC0, so it is not in the critical path. Then, instruction 0xC0 is a conditional branch taken directly from EX stage to instruction 0xD0, instruction 0xA2, 0xC2 and 0xC4 are dropped. This critical path from EX to $fetch\_req$ is to improve the fetch efficiency to save one cycle. However, it limits the frequency.

To cut the first type of critical path, we integrate a 4x32-bit ring FIFO buffer to simplify the cores' instruction fetch (Fig.~\ref{fig:critical_opt}). The small ring FIFO buffer has the following features: 1) the core can send non-blocking fetch requests when the FIFO is not full; 2) the FIFO is empty when the write pointer equals the read pointer; 3) the ring FIFO helps act as a primary cache when the short branches hit the ring FIFO. To cut the second type of critical path from EX for the conditional branch, we must insert a pipeline or implement a branch predictor in the IF/ID stage. However, the branch predictors with branch addresses registering for diverse branches take huge extra area and power. Besides, branch predictors' index searching is also in the critical path to the $fetch\_req$. In the end, we choose to delay the conditional branch one cycle to increase the frequency.

Fig.~\ref{fig:critical_opt} shows the two-level cache after IF optimization, and we can see that a small 4x32-bit ring FIFO buffer is used to issue the $fetch\_req$ without dependence on other signals. Besides, a 128-bit L0 buffer is still used to avoid heavy request traffic to the L1 cache control unit. As a result, it brings more power than the legacy 128-bit IF with the extra ring FIFO buffer. In the end, the final remaining path (red arrow) starts from the L1 caches' DATA array to the cores' inner instruction decompression logic, which is small.

\section{Physical Implementation Results}

This section presents a comprehensive physical exploration of timing, area, and power for all the configurations shown in Table \ref{table:configuration} to characterize the performance and energy efficiency of alternative instruction cache architectures. The comparison includes the three instruction cache variants described in section \ref{sec:background}: private caches (PR), single-port shared caches (SP), and multi-port shared caches (MP), along with four cache configurations proposed in this work. They are the baseline hierarchical cache with prefetcher disabled (HIER) and enabled (HIER\_PRE), and the hierarchical cache with optimized core's pipeline with prefetcher disabled (HIER\_OPT), and enabled (HIER\_OPT\_PRE). Table \ref{table:configuration} also shows the number of cycles required to refill from L2 for the different configurations, which is useful for understanding the results presented in the evaluation sections. In this work, we mainly consider shared and hierarchical caches featuring a size of 4~KB in the last level, providing a good trade-off between performance, area, and efficiency. Furthermore, we consider a cluster with 8 512 B private caches as the baseline. More insight on performance/area trade-offs for different private (PR) and shared cache sizes (SP, MP) can be found in \cite{b1}.

\subsection{Experimental Setup}

To perform a detailed analysis of the trade-offs among the discussed cache architectures in terms of performance, energy, and area, we consider a single cluster with eight cores and an instruction cache with 4~KB (8~KB has the same characteristics as 4~KB as shown in previous work \cite{DATE_Jie}). Besides, the size of the TCDM is fixed to 128~KB. Table \ref{table:configuration} shows all the architectures used in the proposed experiments. A 4-way set associative configuration with a cache line of 16 bytes (4 words in 32-bit) is used, providing a well-balanced trade-off between performance and area overhead for low-power multi-cores \cite{b1}. 

\begin{table}[t]
\begin{center}
\resizebox{\columnwidth}{!}{%
\begin{tabular} { c c c l }
\hline
 Mnemonic & Type & Refill & Description \\
          &      & Cycles &             \\
 \hline
 \textit{PR} & Private & 15 & 512B I\$ bank per core\\
 \hline
 \textit{SP} & Shared & 17 & 8x 512B I\$ banks, single-port\\
 \hline
 \textit{MP} & Shared & 19 & 2x 2048B I\$ banks, 8-port\\
 \hline
 \textit{HIER} & Private & 19 & 512B I\$ bank per core\\
 \cline{2-4}
         & Shared  &    & 2x 2048B I\$ banks, single-port\\
         &         &    & with Response buffer\\
 \hline
 \textit{HIER\_PRE} & Private & 19 & 512B I\$ bank per core with prefetch\\
 \cline{2-4}
             & Shared  &    & 2x 2048B I\$ banks, single-port\\
             &         &    & with Response buffer\\
 \hline
 \textit{HIER\_OPT} & Private & 19 & 512B I\$ bank bank per core\\
 \cline{2-4}
             & Shared  &    & 2x 2048B I\$ banks, single-port\\
             &         &    & with Response buffer\\
 \hline
 \textit{HIER\_PRE\_OPT} & Private & 19 & 512B I\$ bank per core with prefetch\\
  \cline{2-4}
             & Shared  &    & 2x 2048B I\$ banks, single-port\\
             &         &    & with Response buffer\\
\hline
\end{tabular}
}
\end{center}
\caption{4KB Instruction Cache Architectures Explored in this work.}
\label{table:configuration}
\vspace{-0.5cm}
\end{table}

To obtain the physical characteristics of each design (i.e., power, area ), we synthesized and implemented the designs in 22nm FD-SOI technology (full place and route). We used Cadence Genus-19.10 to synthesize the design and Cadence Innovus-18.10 for the place and route. 

To characterize the power consumption, we used the TT, 25$^{\circ}$C, 0.65V corner, commonly used for parallel low-power architectures. First, we simulate the tests with a post-layout netlist in 200Mhz operating frequency (we scale the cluster dynamic power to the actual maximum operating frequency of each architecture). Then, for each test, we back-annotate the switching activity traces in Value Change Dump (VCD) format with Questasim 10.7b only in the third execution to avoid cold boot. Next, we pass it with the parasitic file (in SPEF format) of a specific RC corner to Synopsys PrimePower M-2019.12 to achieve an accurate cluster power estimation. Finally, by considering the L2 memory power, we calculate the total system power and energy.

\subsection{Area and Timing}

\begin{figure}[t!]
\includegraphics[width=0.45\textwidth]{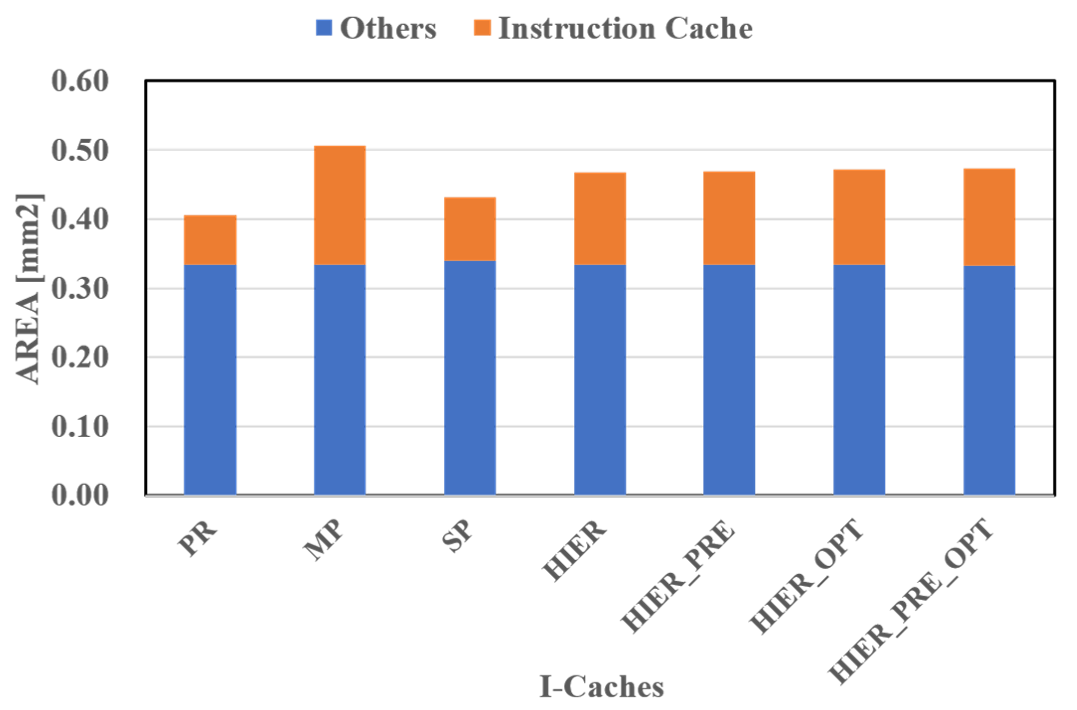}
\caption{Cluster area breakdown for the different cache architectures.}
\label{fig:icache_area}
\vspace{-0.5cm}
\end{figure}

One of the main points for designing a hierarchical cache leads to improved timing behaviour and scalability with the number of cores over the shared cache while preserving the capability to provide large capacity with a small area overhead. Fig.~\ref{fig:icache_area} illustrates the silicon area costs for all cache architecture configurations. Thanks to the private cache's simpler structure, it is smaller than other caches. The two-level cache is bigger than \textit{SP} since it always has more 4~KB (8 x 512 B) L1 memory than \textit{SP}, and \textit{MP} has the largest area because of the multiple reading ports.

The two-level caches feature an area falling between that of \textit{SP} and \textit{MP}. Compared to \textit{SP}, the area increases due to the small private L1 caches, despite the smaller timing pressure on the low-latency interconnect thanks to the decoupled combinational path between the cores and the shared cache banks, reducing its area as well. One interesting point is that according to the proposed prefetch scheme,  the prefetch in a two-level cache brings little additional area, about 1\%, which keeps the same power as a simple two-level cache. Besides, no extra critical path is introduced since the prefetch logic is only inside the L1 cache controller. It is the same for \textit{HIER\_OPT} and \textit{HIER\_PRE\_OPT}, which keep almost the same area.

Table \ref{table:timing} shows the results of the static timing analysis for all the cache architectures, implemented in clusters of 8 cores or 16 cores to highlight the better scalability of the proposed cache. Not surprisingly, the clusters featuring shared caches (\textit{SP}, \textit{MP}) have worse maximum frequency due to the long critical path between the core and the interconnect and the high congestion of the multiple reading ports, respectively. For a system with 8 cores, the caches with simple L1 private cache have the best timing, keeping about 6\% timing improvement compared to shared caches. When we increase the system core number to 16, we observe that \textit{MP} and \textit{SP} have about 19\% and 15\% maximum frequency drop compared to \textit{HIER\_OPT}, respectively. For the \textit{MP}, 16-port memory banks cause serious wire congestion, leading to worse timing results. The same issue arises for \textit{SP}: increasing the number of channels also increases the levels of logarithmic interconnect, and the critical paths become worse due to the combinational dependency between the \textit{request} and \textit{response} channels. Thus, the two-level cache with a non-blocking fetch interface has better timing and scalability by eliminating the long paths in the \textit{request} and \textit{response} channels.



\begin{table}[t!]
\begin{center}
\resizebox{\columnwidth}{!}{%
\begin{tabular} { c c c c c c c }
\hline
Type &  \multicolumn{2}{c}{Max frequency} & \multicolumn{2}{c}{Speed up vs. \textit{PR}}  \\
\cline{2-5}
     &      8-core     &    16-core       &  8-core     &    16-core\\
\hline
 \textit{PR}             & 378 MHz  & 363 MHz &   -     & -       &\\
 \textit{MP}             & 357 MHz  & 306 MHz &  -5.6\% & -15.7\% &\\
 \textit{SP}             & 350 MHz  & 320 MHz &  -7.4\% & -11.8\% &\\
 \textit{HIER}           & 372 MHz  & 354 MHz &  -1.6\% & -2.5\%  &\\
 \textit{HIER\_PRE}      & 372 MHz  & 354 MHz &  -1.6\% & -2.5\%  &\\
 \textit{HIER\_OPT}      & 429 MHz  & 399 MHz &  +13.5\% & +9.9\% &\\
 \textit{HIER\_PRE\_OPT} & 429 MHz  & 399 MHz &  +13.5\% & +9.9\% &\\
\hline
\end{tabular}
}
\end{center}
\caption{Results of the Static Timing Analysis.}
\label{table:timing}
\vspace{-0.5cm}
\end{table}

\subsection{Parallel Performance and Power Characterization}
\label{sec:power_model}

This section presents a detailed performance and power characterization of the presented instruction cache architectures to provide a direct comparison between the different solutions as well as to create an analytical model for power estimation of real-life applications.

Obtaining the power of the whole cluster system with different instruction caches for real-life applications running for millions of clock cycles on a post place and route database is not feasible, both due to the long simulation time and size of the VCD traces required to annotate the switching activity of the design. Moreover, real-life applications often have complex and unpredictable behaviours, making it difficult to understand the functional and power trends of the cache. To model the presented caches and provide more insight into their behaviour regarding power (and energy) consumption, we propose a methodology based on a synthetic benchmark where we artificially modulate the instruction locality. 

We created a synthetic test performing a parallel vector multiplication between 8192 elements, we distribute the workload evenly over $TOTAL\_CORE$ number of cores, and we use loop unrolling to control the size of the batch of instructions executed by the processors (controlled through variable $STEP$, as shown in algorithm \ref{alg:loop_unrolling}. The $BUFFER\_SIZE$ is fixed to 8192 while $STEP$ changes among 32, 64, 128, 256, 512, and 1024, producing loops iterating over a loop body of 0.375~KB, 0.75~KB, 1.5~KB, 3~KB, 6~KB, and 12~KB, respectively. Since the applications executed by the cores have an IPC close to 1, the only stalls of the cores are those caused by cache misses, decreasing the whole cluster system's activity and reducing its power consumption. On the other hand, the power consumption of the L2 memory increases with the miss rate due to the increasing number of refills. To take into account this significant contribution, we characterized the energy consumption of every refill from L2 and the overall L2 leakage power and added them to the system power. We used this methodology to characterize the behaviour of the seven architectures, summarized in equation \ref{eq1}. We obtain the parameters $L2\_leakage\_power$ and $L2\_per\_read\_energy$ from the L2 SRAM's datasheet. The parameters $cycles$ and $L2\_miss\_refill\_number$ can be read from the hardware counters implemented inside the instruction cache with the cycle-accurate simulation. With a specific system frequency, we only need to find the $cluster\_power$ to have the total energy.

\begin{footnotesize}
\begin{equation}\label{eq1}
\begin{aligned}
& Total\_energy \\
& = (cluster\_power + L2\_leakage\_power) \cdot time + L2\_read\_energy  \\
& = (cluster\_power + L2\_leakage\_power) \cdot cycles / frequency \\
& + L2\_miss\_refill\_number \cdot L2\_per\_read\_energy
\end{aligned}
\end{equation}
\end{footnotesize}

\begin{algorithm}[t!]
\caption{Parallel Synthetic Benchmark}
\label{alg:loop_unrolling}
\begin{algorithmic}
\Require $BUFFER\_SIZE = 8192$
\Require $TOTAL\_CORE  = 8$
\Require $STEP         = 32$ \\

\State $start \gets core\_id \times BUFFER\_SIZE \  / \ TOTAL\_CORE$
\State $end   \gets start + BUFFER\_SIZE \ / \ TOTAL\_CORE$\\

\For{$i = start; i < end; i += STEP $}
    \State $c[i] \gets a[i] \times b[i]$
    \State $c[i] \gets a[i+1] \times b[i+1]$
    \State $...$
    \State $c[i+STEP-1] \gets a[i+STEP-1] \times b[i+STEP-1] $
\EndFor

\end{algorithmic}
\end{algorithm}

Summarized performance, power, and energy results are presented in the following. Results are normalized to the 8-core \textit{PR} configuration using the version of the synthetic benchmark featuring a loop body as large as 0.375~KB, which we consider as a baseline for our exploration. In all the experiments, the caches are warmed-up before measuring the throughput.

\begin{figure*}[t!]
\subfloat[With fixed frequency, 200MHz \label{fig:test_result_throughput}]{
  \includegraphics[width=0.49\textwidth]{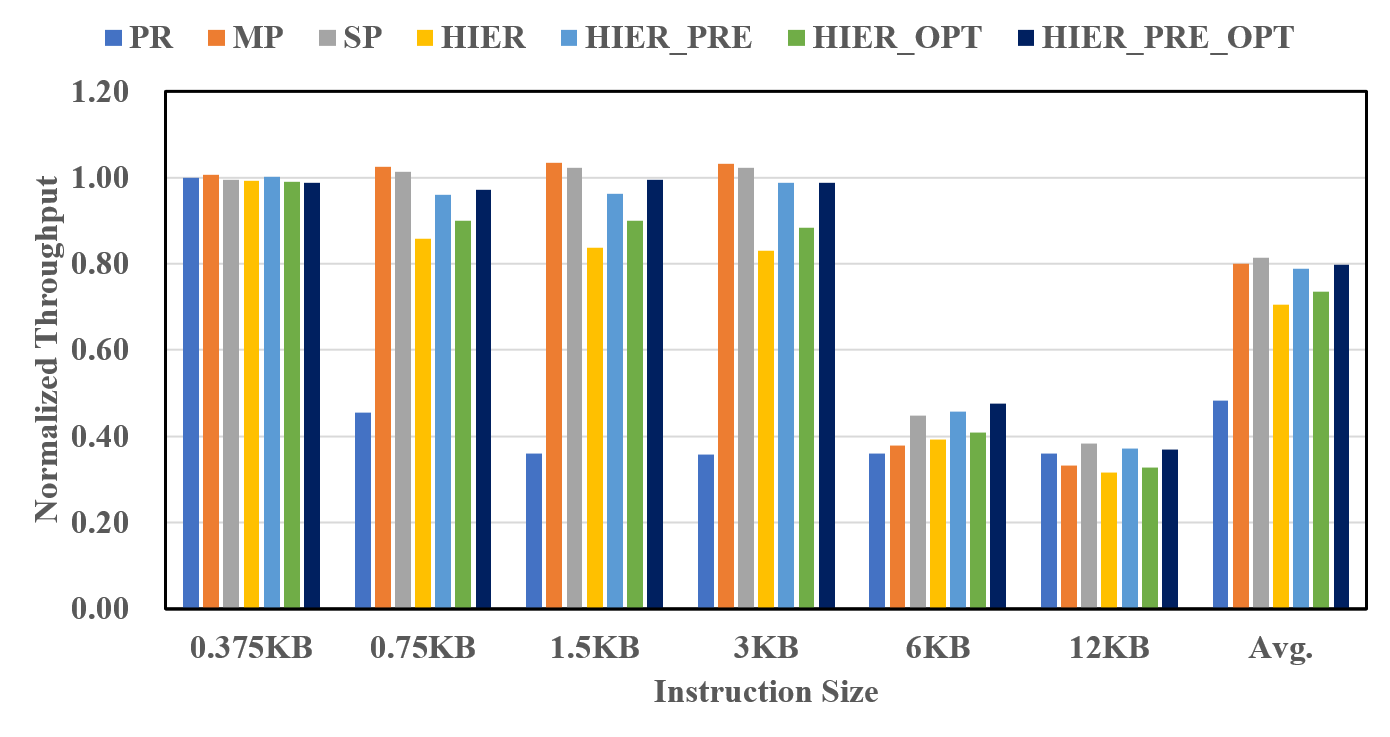}%
}
\hfill 
\subfloat[With maximum frequency\label{fig:test_result_throughput_max}]{
  \includegraphics[width=0.49\textwidth]{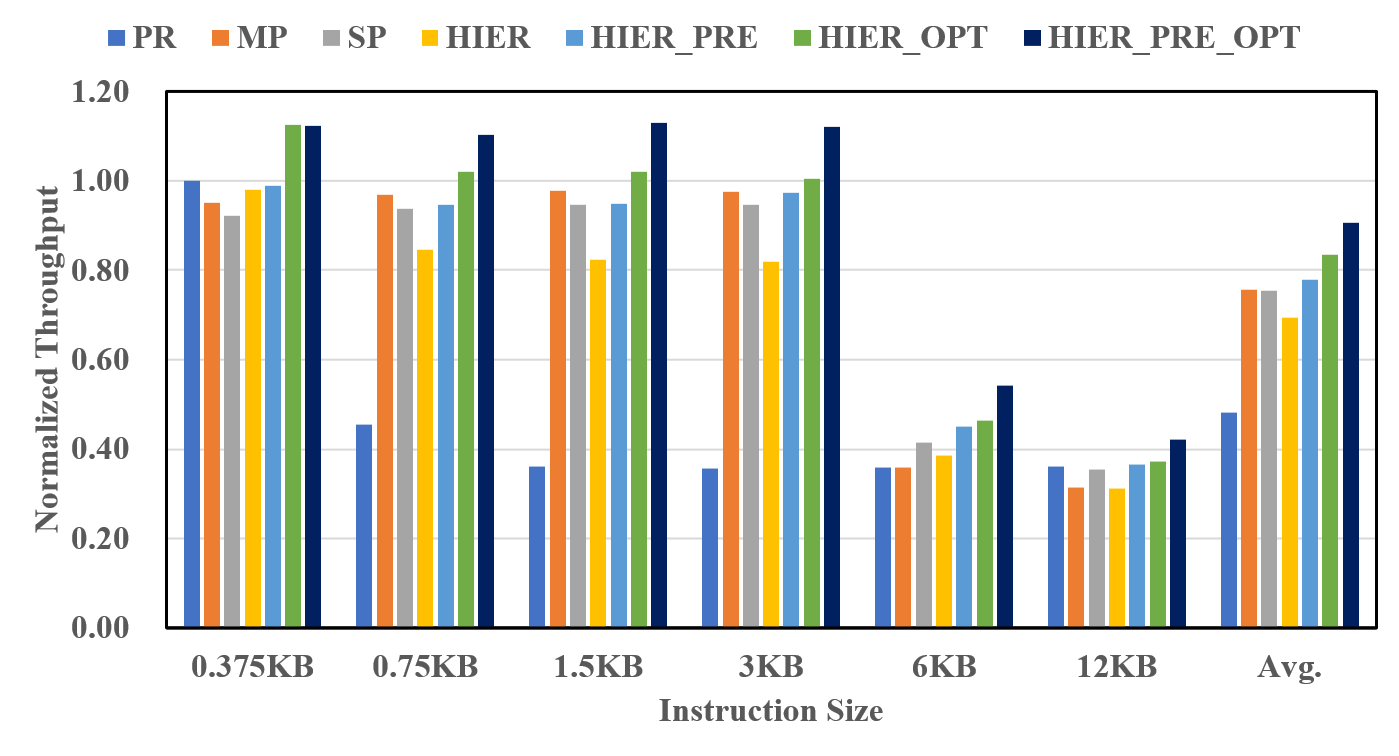}%
}
\caption{Throughput of tests normalized to \textit{PR} with 0.375~KB cache size.}
\label{fig:test_result_per}
\vspace{-0.5cm}
\end{figure*}

\begin{figure*}[t!]
\subfloat[Power\label{fig:test_result_power}]{%
  \includegraphics[width=0.5\textwidth]{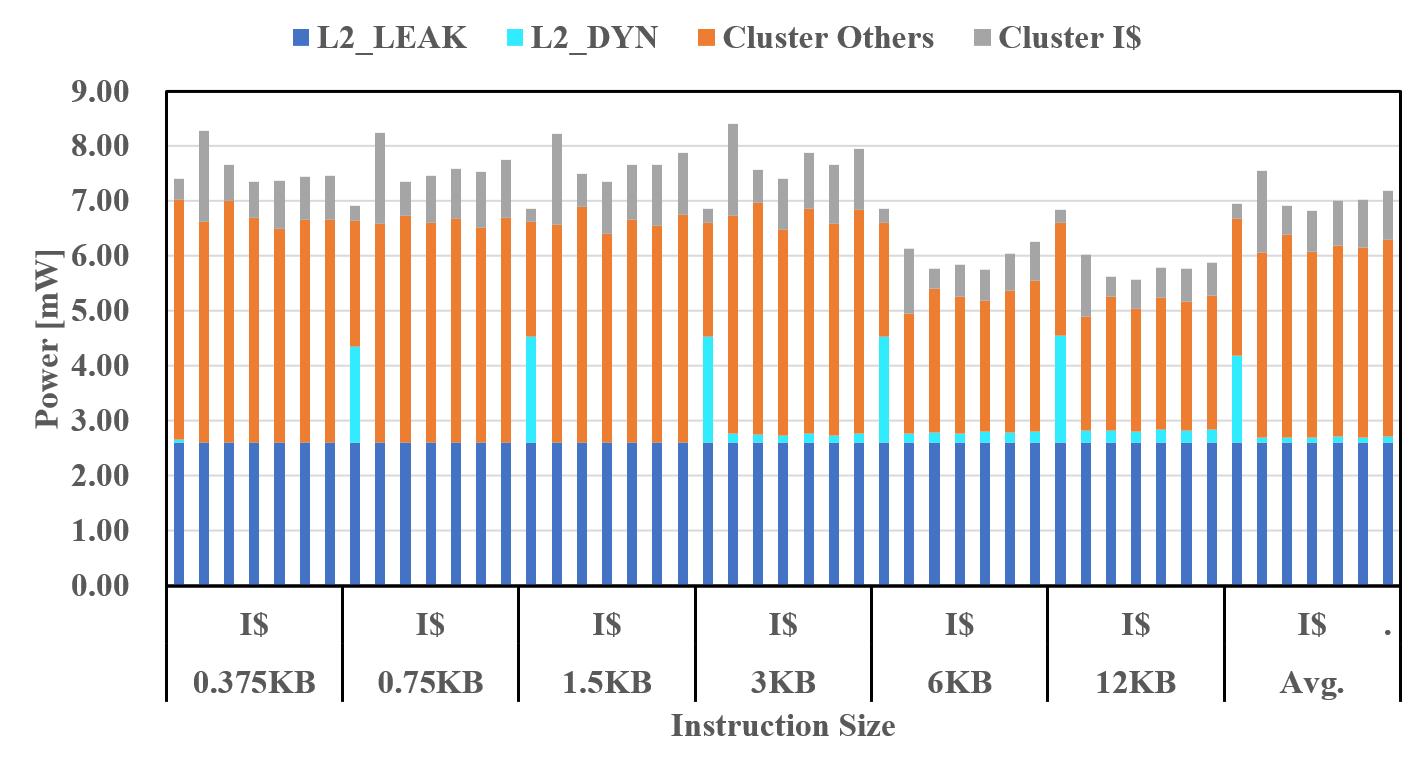}%
}
\hfill
\subfloat[Energy efficiency\label{fig:test_result_energy}]{%
  \includegraphics[width=0.5\textwidth]{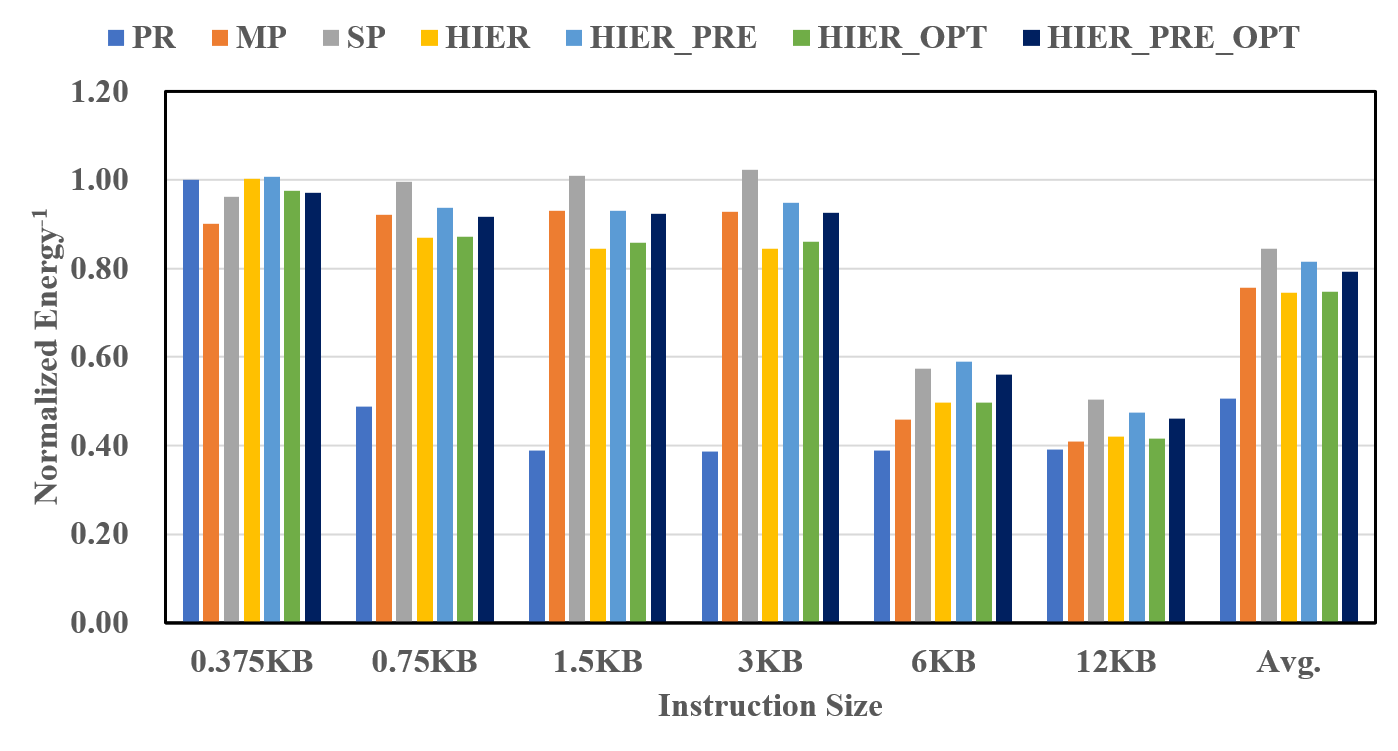}%
}
\caption{Power and energy efficiency of tests normalized to \textit{PR} with 0.375~KB cache size, 200MHz. Each \textit{I\$} in Fig. (a) represents different I-Caches. \textit{I\$} = \{\textit{PR MP SP HIER HIER\_PRE HIER\_OPT HIER\_PRE\_OPT}\} }
\label{fig:synth_test}
\vspace{-0.5cm}
\end{figure*}

Fig.~\ref{fig:test_result_throughput} shows the normalized throughput of the synthetic tests running on the different architecture configurations, assuming that all the configurations are running at the same operating frequency, providing an insight into their functional performance. When the loop body is smaller than the size of L1 of the HIER caches (0.5~KB), the throughput of all the configurations is the same since they always hit in L1. When increasing the size of the loop body to 0.75~KB, the performance of the PR cache drops significantly ($\sim$55\%) since each cache miss is refilled from L2, implying 15 cycles of latency. On the other hand, the HIER configurations feature only a slight drop in performance ($\sim$15\%) thanks to the low latency of the refills from the 4~KB L1.5, which is almost completely recovered when activating the prefetcher. When the loop body is larger than 4~KB (size of shared configuration and L1.5 of the HIER configuration), the performance of these configurations also starts dropping due to capacity miss. 

Fig.~\ref{fig:test_result_throughput_max} shows the normalized throughput when each configuration run at the maximum operating frequency. It is possible to note that \textit{HIER\_PRE\_OPT} improves the performance by $\sim$16\% for all the benchmarks with respect to shared caches, thanks to the similar functional performance and much higher maximum operating frequency. In particular, most of the gain is thanks to the optimized instruction fetch stage of the RI5CY core described in section \ref{if_opt}.

Fig.~\ref{fig:test_result_power} shows the power of all synthetic tests measured on different architecture configurations. The \textit{PR} configuration is the one consuming less power in all configurations, thanks to the simple and straightforward implementation for loop bodies below 0.5~KB; for larger loop bodies, the power of the \textit{PR} configuration drops due to the high miss rate. In general, SP is the configuration featuring the smallest power consumption. However, we note that for the test with the 0.375~KB loop body (smaller than the L1 cache capacity), \textit{HIER\_PRE} still has less power than \textit{SP}. Despite the better functional performance, the prefetcher brings additional power consumption. This is evident by looking at \textit{HIER\_PRE} configurations, consuming 3\% more power than \textit{HIER} on average. Besides, \textit{HIER\_OPT} and \textit{HIER\_PRE\_OPT} have one more ring buffer FIFO described in section \ref{if_opt} than \textit{HIER}, so they bring 2\% and 3\% more power than \textit{HIER} on average, respectively.

Fig.~\ref{fig:test_result_energy} shows the energy efficiency ($ Energy^{-1}$) of all synthetic tests measured on the different architecture configurations. In general, the \textit{SP} is the cache providing the best trade-off between performance and power, delivering better energy efficiency on these synthetic benchmarks. Besides, two-level caches are penalized due to the higher power consumption caused by their two-level nature. The versions without a prefetcher are further penalized by the worse functional performance, which is recovered by enabling the prefetcher to bring 10\% to 20\% better performance with only $\sim$3\% increase of power.




To validate the power model, we analyze the error rate between LUT and exhaustive power estimation on three of the smallest real-life applications considered in this work. The previous work \cite{DATE_Jie} selected several relatively small applications and estimated the error rate between power estimation using PrimeTime and LUT evaluation model, with a maximum error smaller than 6\%. Since we only care about the relationship of each cache, so all power results are normalized to PR. Finally, we use this accurate power estimation to make a detailed comparison. This model poses the basis for evaluating larger real-life applications that cannot be executed in a reasonable time on post-layout netlists.
\section{Benchmarking}

\subsection{Description of Parallel Benchmarks}

To analyze in-depth the behaviour of the presented instruction cache architecture, we use a set of benchmarks relevant for several application domains in the market of IoT processors \cite{GAP9} such as image, audio, and vibration processing. The applications are based on full-fledged optimized OpenMP \cite{b17} applications and four CNN applications \cite{b18}, each one featuring a different behaviour in terms of access patterns to the instruction memory subsystem, as well as diversified memory footprints and execution time.

Table \ref{table:application} shows the detailed characteristics of each application, including the code size and the number of instructions executed. Since cache performance is influenced strongly by code locality and size, we classify the applications into three groups. The short-Jump class includes BFS and CNN kernel-based applications, which are loop-based applications with most loop bodies smaller than four lines of cache. The second class, long-Jump, groups all the loop-based applications with loop bodies greater than four cache lines or based on extensive control flow instructions, including CT, FAST, and SLIC. Finally, the library includes HOG, SRAD, and FFT. These kernels use support libraries to manage floating-point emulation and fixed-point arithmetic, with the specific purpose of generating significant stress in the cache \cite{b1}. In the IoT domain, these functions are typically inlined or supported in hardware; however, we included them in the benchmark set to put an upper bound on average recurrency and size of program counter's jumps.

The above applications are fairly complex and long-lived (millions of instructions in most cases), so it takes too much time to explore with RTL simulation. For this reason, we analyze the performance based on measures coming from RTL-equivalent FPGA implementations mapped on Xilinx Zynq ZCU102 FPGA using Vivado 2019.2. The FPGA emulation allows executing at up to 50 MHz, enabling near-to-real-life execution time. The performance analysis is based on statistics collected by hardware counters implemented inside the instruction cache. Then we calculate the miss rate of each application and use the power model described in section \ref{sec:power_model} to estimate the absolute power and energy efficiency for each cache architecture.

\begin{table}
\begin{center}
\resizebox{\columnwidth}{!}{%
\begin{tabular} { c c c l }
 \hline
 APP & Size [KB] &  Class & Description \\
 \hline
 BFS  & 59.2 &  Short-Jump & Breadth-First Search\\
 CT   & 28.2 &  Long-Jump & Color Tracking\\
 FAST & 28.6 &  Long-Jump & Machine-generated corner Detection \\
 SLIC & 26.1 &  Long-Jump & Simple Linear Iterative Clustering\\
 HOG  & 33.3 &  Library & Histogram of Oriented Gradients\\
 SRAD & 31.6 &  Library & Speckle Reducing Anisotropic Diffusion\\
 FFT  & 41.2 &  Library & Fast Fourier transform \\ 
 CIFAR10   & 37.5 &  CNN & Object Recognition\\
 MNIST     & 37.9 &  CNN & Handwritten digits Recognition\\
 KWS       & 30.1 &  CNN & Keyword spotting\\
 CNNDronet & 71.3 &  CNN & Detector for Real-Time UAV Applications \\
 \hline
\end{tabular}
}
\end{center}
\caption{Parallel benchmark details}
\label{table:application}
\end{table}

\begin{table} [tp!]
\begin{center}
\resizebox{\columnwidth}{!}{%
\begin{tabular} { c c c c c c c c c c c c c }
\hline
     & \%
     & \textit{PR} & \textit{MP} & \textit{SP} 
     & \multicolumn{2}{c}{\textit{HIER}}
     & \multicolumn{2}{c}{\textit{HIER\_PRE}}
     & \multicolumn{2}{c}{\textit{HIER\_OPT}}
     & \multicolumn{2}{c}{\textit{HIER\_PRE\_OPT}}\\
     &   &     &      &        &  L1 & L1.5  &  L1 & L1.5 & L1 & L1.5 & L1 & L1.5      \\

\hline
Short       & BFS       &  0.5&	0.0&	0.0&	0.5&	0.0&	0.2&	0.0 & 0.6 &0.0 & 0.1 &0.0 \\
\cline{1-13}
CNN         & CIFAR10   & 2.8&	0.1&	0.2&	2.8&	2.8&	0.9&	2.1&   2.9&	2.6 & 0.5 & 2.5 \\
            & MNIST     & 3.8 &	0.0 &	0.0&	3.8&	0.38&	1.45&	0.36&	3.9 &	0.5 & 0.7 &0.5 \\
            & KWS       & 1.2&	0.0&	0.0&	1.2&	0.1&	0.6&	0.0&	1.29 &	0.05 & 0.37 &0.02\\
            & DRONET    & 9.3&	0.0&	0.1&	9.3&	0.2&	3.3&	0.2&	9.4 &	0.2 & 3.2 & 0.2\\
\cline{1-13}
Long        & CT        & 0.6&	0.0&	0.0&	0.6&	0.2&	0.2&	0.0&   0.7 &	0.3 & 0.2 &0.1\\
            & FAST      & 5.8&	0.1&	0.4&	6.2&	2.1&	2.3&	1.3&	5.5 &	1.8 & 1.6 &2.3\\
            & SLIC      & 0.5&	0.0&	0.0&	0.5&	1.5&	0.2&	1.2&	0.6 &	1.8 & 0.2 & 0.7\\
\cline{1-13}
Library     & HOG   & 20.0&	0.0&	0.1& 19.5&	0.1&	            
            6.2$\downarrow$&	0.1& 20.3&	0.1 & 
            6.5$\downarrow$ &0.1\\
            & FFT    & 54.8&	0.1&	0.3&	54.7&	0.2&	30.7$\downarrow$&	0.1&	59.8&	0.2 & 26.7$\downarrow$ &0.1\\
            & SRAD   & 54.5&	0.1&	0.5&	54.3&	0.2&	27.5$\downarrow$&	0.2&	57.6&	0.2 & 20.7$\downarrow$ &0.2\\
\cline{1-10}
\hline
\end{tabular}
}
\end{center}
\caption{Miss rate of real-life applications. The L1 prefetch decreases the L1 miss rate significantly.}
\label{table:app_miss}
\vspace{-0.5cm}
\end{table}


\begin{figure*}[tp!]
\subfloat[Low L1 miss rate applications.]{%
  \includegraphics[width=0.49\textwidth]{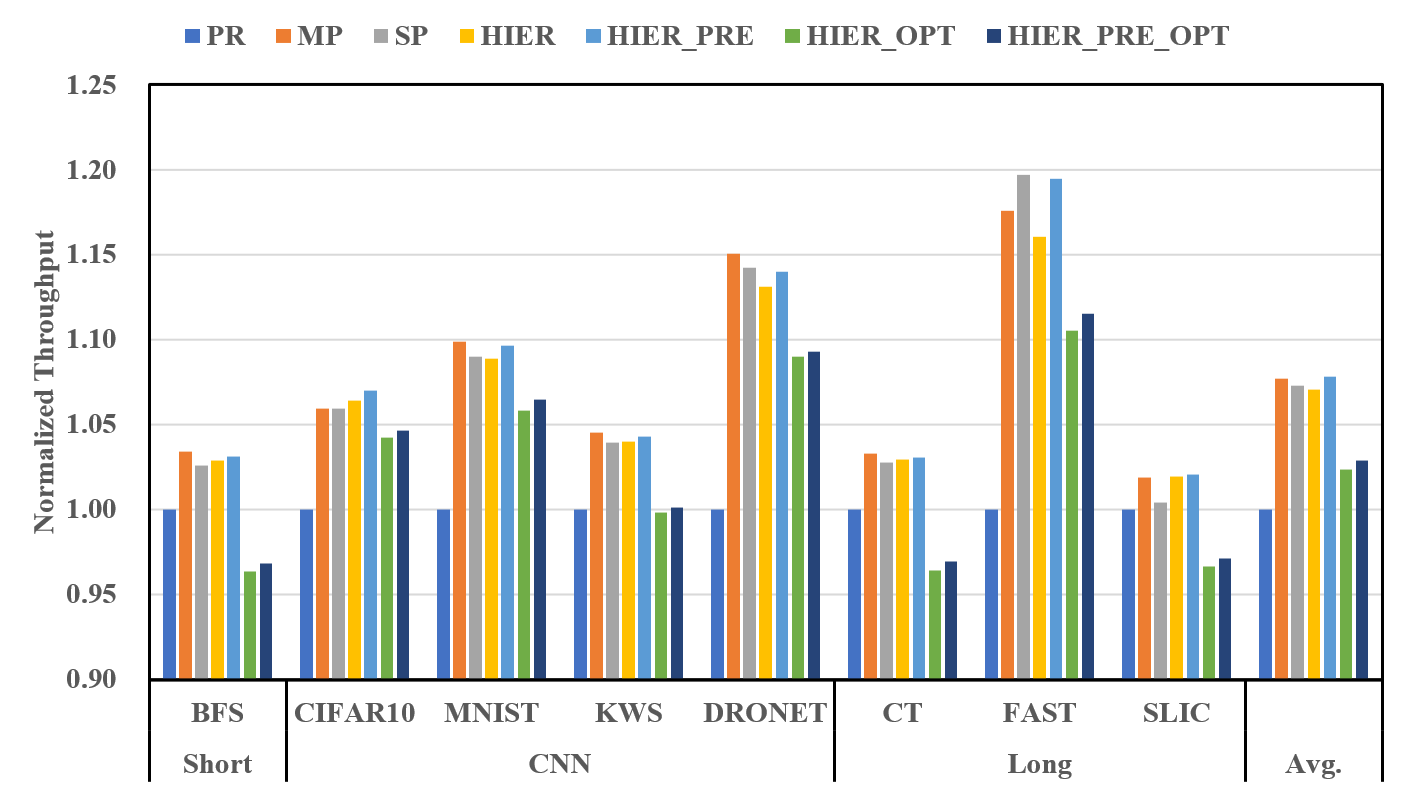}%
}
\subfloat[High L1 miss rate applications.]{%
  \includegraphics[width=0.49\textwidth]{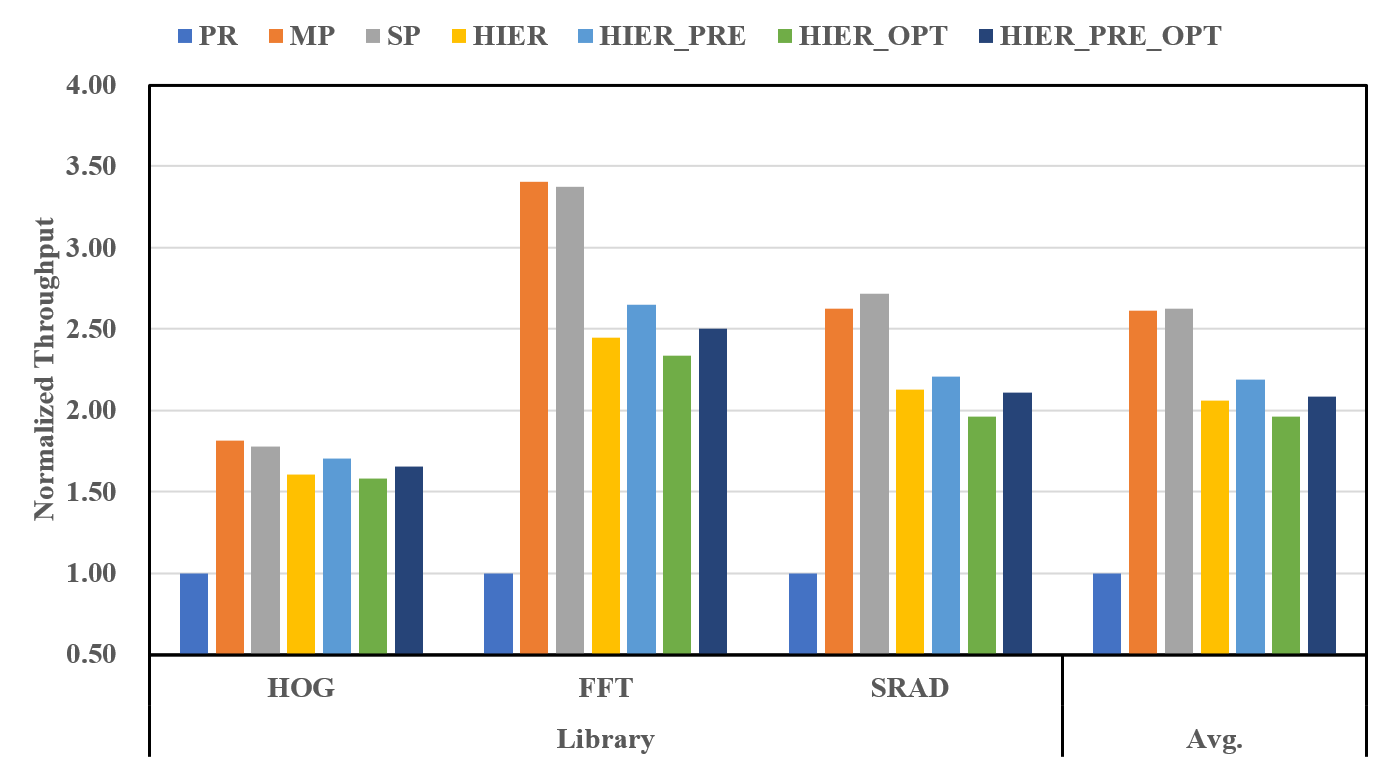}%
}
\caption{Throughput of real-life applications normalized to \textit{PR}, 200MHz.}
\label{fig:app_perf}
\vspace{-0.7cm}
\end{figure*}

\begin{figure*}[tp!]
\subfloat[Low L1 miss rate applications.]{%
  \includegraphics[width=0.5\textwidth]{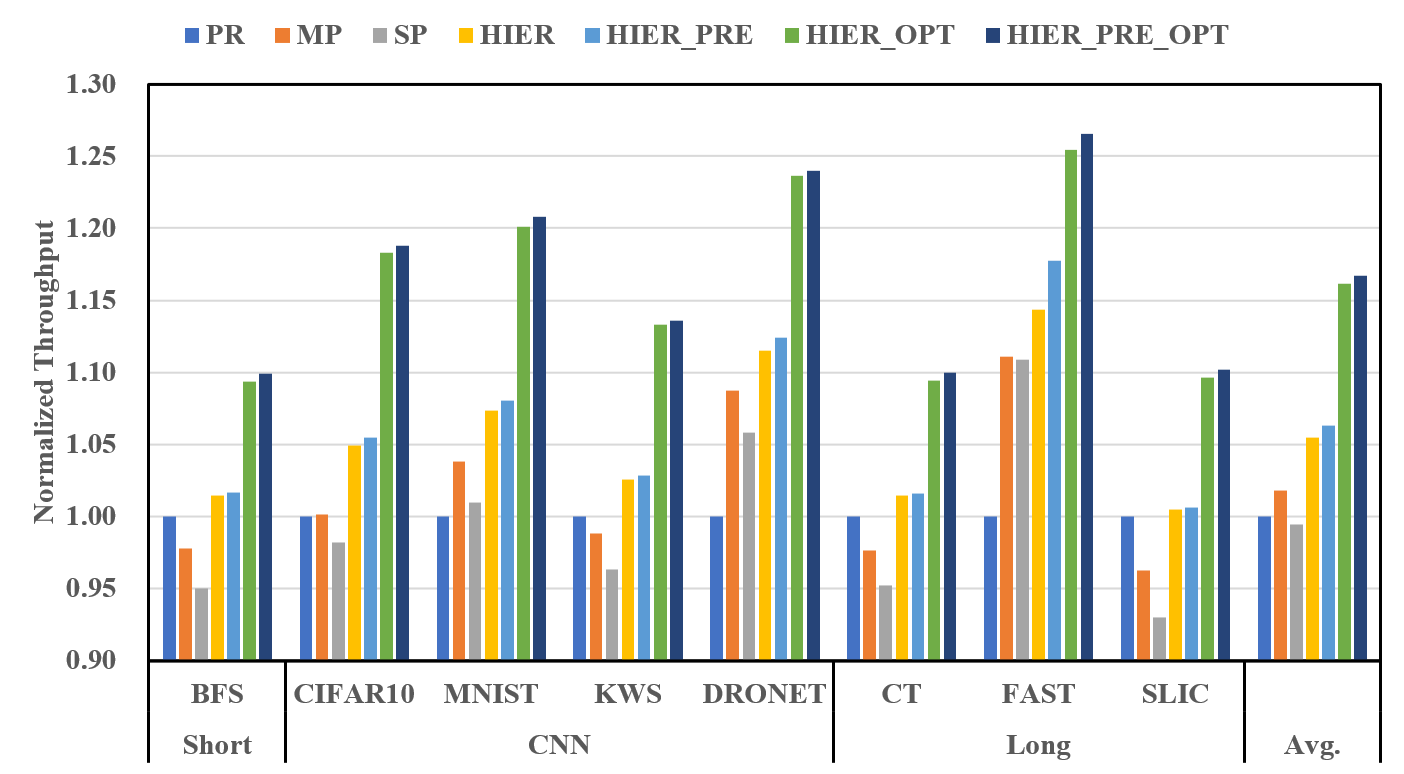}%
}
\subfloat[High L1 miss rate applications.]{%
  \includegraphics[width=0.5\textwidth]{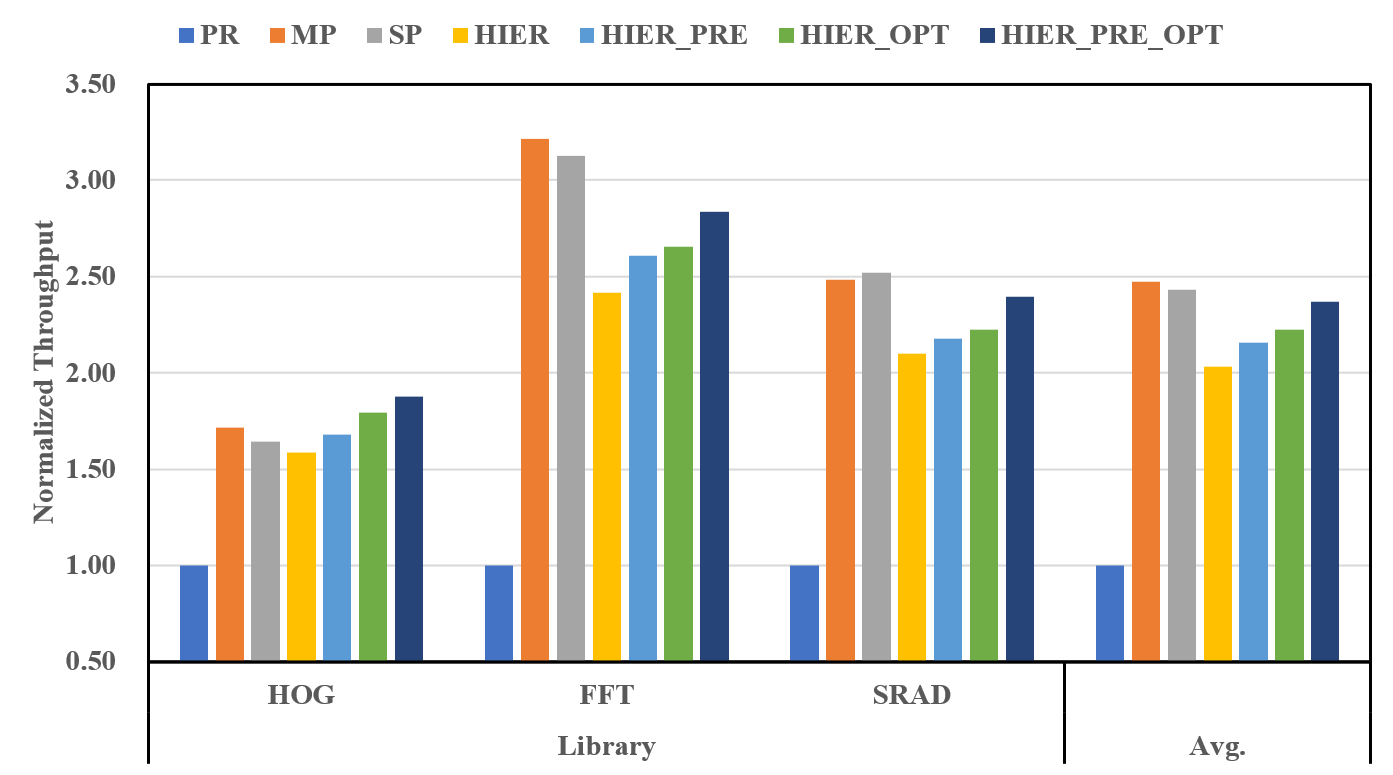}%
}
\caption{Throughput of real-life applications normalized to \textit{PR} at maximum frequency.}
\label{fig:app_perf_max}
\vspace{-0.7cm}
\end{figure*}

\begin{figure*}[tp!]
\subfloat[Low L1 miss rate applications.]{
  \includegraphics[width=0.5\textwidth]{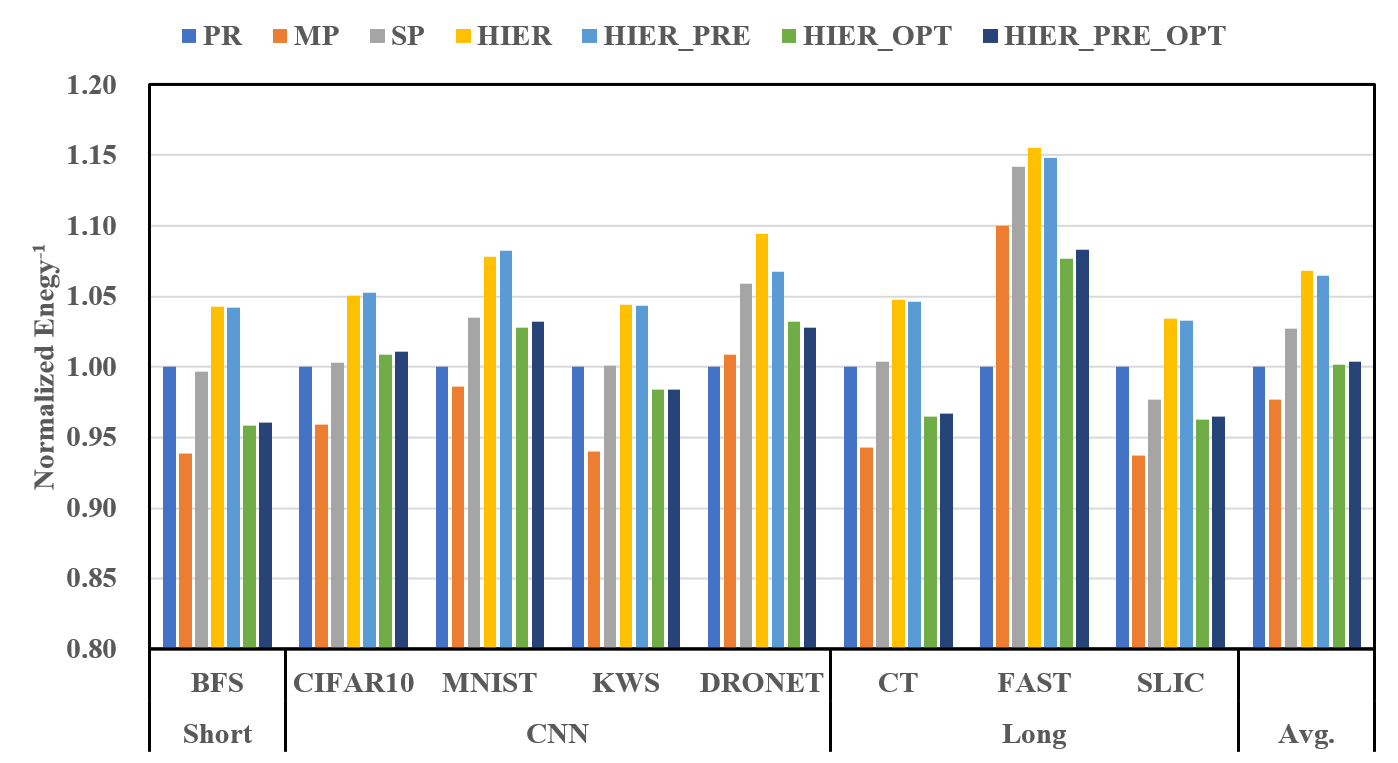}%
}
\subfloat[High L1 miss rate applications.]{
  \includegraphics[width=0.49\textwidth]{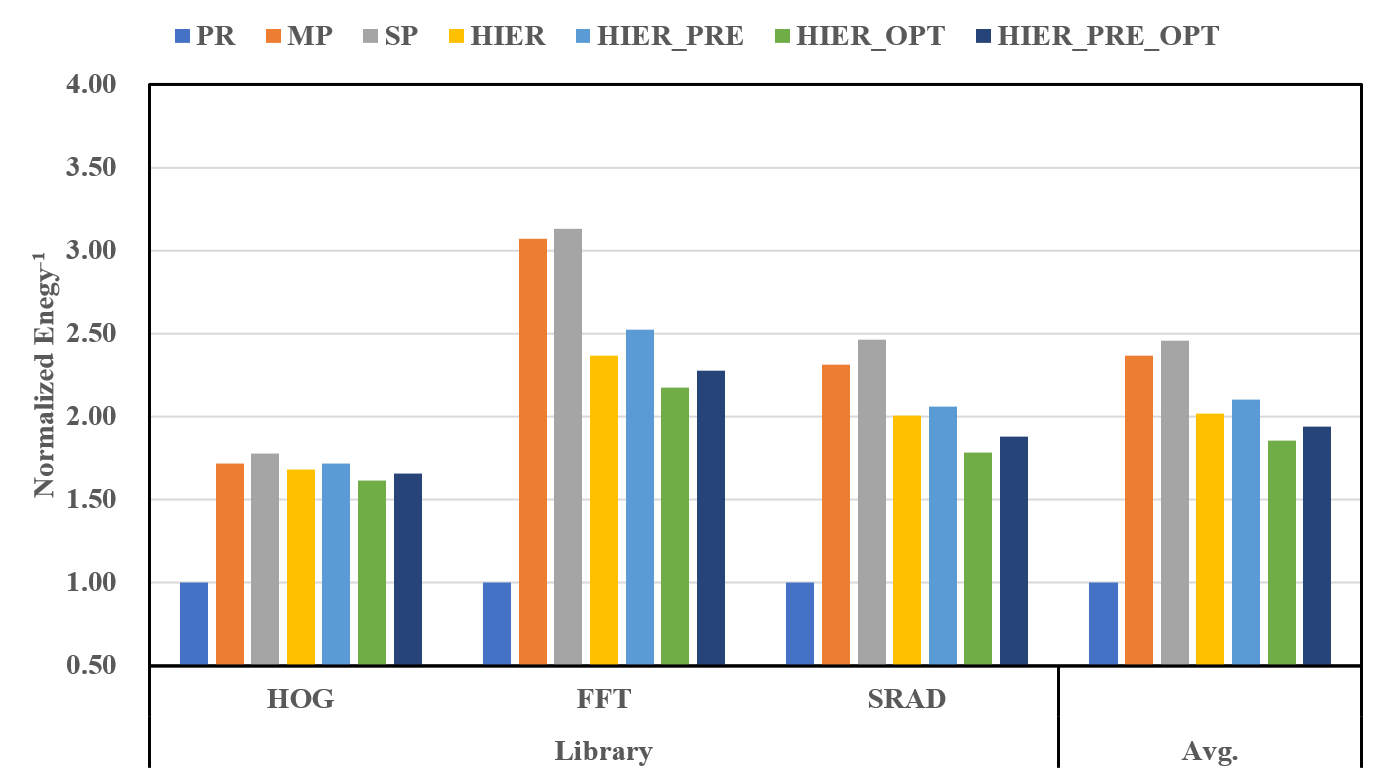}%
}
.\caption{Energy efficiency of real-life applications normalized to PR, 200MHz}
\label{fig:app_energy}
\vspace{-0.5cm}
\end{figure*}

\subsection{Parallel Performance and Power Benchmarking}

Table \ref{table:app_miss} shows the miss rate of real-life applications for the considered cache architectures. \textit{PR}, \textit{SP}, and \textit{MP} only report L1 cache misses, while the two-level cache reports L1 miss rate and L1.5 miss rate. It can be noted that while the \textit{PR} and the L1 of \textit{HIER} feature a reasonable miss rate for short jump, CNN, and long jumps applications, they suffer a significant miss rate when running library-based applications. Not surprisingly, the two-level caches (SP and MP) always feature the lowest miss rate thanks to their large capacity (4kB). On the other hand, it can be noted that the 4kB L1.5 of the \textit{HIER} cache significantly reduces the miss rate thanks to its higher capacity. Finally, the impact of prefetching is especially relevant for library-based applications. Indeed, as highlighted by the arrows in \ref{table:app_miss} for library-based applications, \textit{HIER\_PRE} and \textit{HIER\_PRE\_OPT} reduce the L1 miss rate by about 50\% with respect to hierarchical caches without prefetching. In the following, to ease the comparison, we separate the applications into two groups according to L1 miss rate: the high L1 miss rate applications (library applications) and the low L1 miss rate applications ( CNN, Short-Jump, Long-Jump).

Fig.~\ref{fig:app_perf} shows the functional performance of the proposed cache architectures running at the same operating frequency and normalized to \textit{PR}. In general, since shared caches and two-level caches can remove capacity miss with relatively larger cache size, they always have better performance than PR. Besides, \textit{HIER} has two more cycles refill when L1 miss and L1.5 hit than \textit{SP}, so it loses 2\% performance compared with shared caches. When we compare the performance of the low L1 miss rate applications, \textit{HIER\_PRE} can achieve the same performance with shared caches and always improve performance compared with \textit{HIER}. \textit{HIER\_PRE\_OPT} functional performance is, on average, smaller by 5\% with respect to \textit{HIER} and shared caches due to the stall caused by the pipeline stage added to unconditional branches to improve the operating frequency of the design. For always hit applications such as BFS, CT, and SLIC, the functional performance drop is 3.5\% smaller than PR, on average. Finally, for the high L1 miss rate applications, even though \textit{HIER\_PRE} and \textit{HIER\_PRE\_OPT} reduce the L1 capacity miss and improve the performance by 7\% on average compared with \textit{HIER} and \textit{HIER\_OPT}, their functional performance is about 20\% smaller compared to shared caches mainly due to the additional latency required to refill from L2.

Fig.~\ref{fig:app_perf_max} shows the performance results of the different cache architectures when operating at the maximum operating frequency, highlighting the better scalability of the hierarchical solutions, especially the one with optimized IF. Results are normalized to the performance of \textit{PR}. The \textit{HIER\_PRE\_OPT} improves the performance by 17\% compared with private cache and shared cache for the low L1 miss rate applications. For the high L1 miss rate applications (which are again not common in the IoT domain), \textit{HIER\_PRE\_OPT} delivers 2.4x better performance than the private caches and only about 5\% smaller performance than shared caches.

Fig.~\ref{fig:app_energy} shows the results of energy efficiency with fixed frequency, which are normalized to \textit{PR}, as well. For the low L1 miss rate applications (Fig.~\ref{fig:app_energy}(a)), the improvement of efficiency of \textit{HIER} over \textit{PR} is limited, since the L1 cache suffers less capacity miss. Moreover, the additional power brought by the prefetcher in \textit{HIER\_PRE} is more or less equivalent to the improvement of performance, which keeps the same energy efficiency compared with \textit{HIER}. The same applies when we compare \textit{HIER\_PRE\_OPT} with \textit{HIER\_OPT}. \textit{HIER\_OPT} and \textit{HIER\_PRE\_OPT} consume more power than \textit{HIER} because of the additional 4x32-bit ring FIFO buffer to cut the critical path. Since most of real-life IoT applications have a low L1 miss rate, \textit{HIER}'s small L1 cache takes advantage of lower power and relatively high energy efficiency without losing much performance compared with shared caches (Fig.~\ref{fig:app_energy} (a)). As a result, \textit{HIER} and \textit{HIER\_PRE} feature, on average, 7\% better energy efficiency than all other caches. Besides, the decision to use the prefetch feature in a two-level cache should be considered by software for the low L1 miss applications. For the high L1 miss rate applications (Fig.~\ref{fig:app_energy}(b)), \textit{HIER\_PRE} has the same energy efficiency as \textit{HIER}, which means the additional power gain is equal to the performance gain. Finally, \textit{HIER\_PRE} brings a 7\% improvement in performance while keeping the same energy efficiency compared with \textit{HIER}. Nevertheless, shared caches remove capacity miss with large L1, so they have about 20\% gain in energy efficiency compared with two-level caches. However, since \textit{HIER\_OPT} isolates the critical path from the cores to instruction caches, it brings better scalability for multi-core systems regarding the number of cores per cluster and maximum operating frequency.

\begin{table}[t!]
\begin{center}
\resizebox{\columnwidth}{!}{%
\begin{tabular} { c c c c c c c c c c c l}
\hline
Type & A & \multicolumn{2}{c}{Max Frequency} & \multicolumn{3}{c}{Low L1 miss rate} & \multicolumn{3}{c}{High L1 miss rate} \\
\cline{3-10}
     && 8-core & 16-core & MxP & P & 1/E 
     & MxP & P & 1/E  \\
 \hline
 \textit{PR}            & 1.00 & 1.00 & 1.00 & 1.00   & 1.00  & 1.00 & 1.00 & 1.00 & 1.00 \\
 
 \textit{MP}            & 1.25 & 0.95 & 0.84 & 1.02  & 1.10  & 0.98  & 2.47 & 1.10 & 2.38  \\
 
 \textit{SP}            & 1.06 & 0.93 & 0.88 & 0.99  & 1.04  & 1.03  & 2.43 & 1.06 & 2.48  \\
 
 \textit{HIER}          & 1.17 & 0.99 & 0.98 & 1.06  & 1.00  & 1.07  & 2.03 & 1.01 & 1.99  \\
 
 \textit{HIER\_PRE}     & 1.17 & 0.99 & 0.98 & 1.06  & 1.01  & 1.07  & 2.16 & 1.04 & 2.11  \\
 
 \textit{HIER\_OPT}      & 1.18 & 1.14 & 1.10 & 1.16  & 1.01  & 1.00 & 2.22 & 1.05 & 1.87 \\
 
 \textit{HIER\_PRE\_OPT} & 1.18 & 1.14 & 1.10 & 1.17  & 1.02  & 1.00 & 2.37 & 1.07 & 1.95 \\
\hline
\end{tabular}
}
\end{center}
\caption{Summary of Area (A), Maximum frequency and Maximum Performance (MxP), Power (P), and Energy Efficiency (1/E) of the instruction caches normalized to \textit{PR}.}
\label{table:summary}
\vspace{-0.5cm}
\end{table}

\subsection{Discussion}

To put the experimental results in perspective, we collect them in Table~\ref{table:summary}. We separate the results into two groups, the low and the high L1 miss rate applications(i.e., Library-based). For the high L1 miss rate applications, which are not common in the IoT domain, both shared and two-level caches feature 2$\times$ better performance than private cache thanks to the large cache capacity. Single-port shared cache features the best energy efficiency, and multi-port shared cache has the maximum performance. Still, the two-level cache performance and energy efficiency are not so far from that of shared caches, and the prefetcher can mitigate the performance drop by reducing it to 5\% at the cost of some more power.

For the low L1 miss rate applications, we note that the two-level cache with optimized instruction fetch subsystem delivers significant maximum performance, up to 17\% larger than private and shared caches. This large advantage in performance is mainly achieved thanks to its higher operating frequency with respect to other solutions. However, this performance increase comes at the cost of an additional area with respect to private caches. While in shared solutions the area overhead is caused by high timing pressure on the interconnect, and multiple read ports and routing congestion for SP and MP, respectively, for the hierarchical cache the overhead in area is structural, due to its 2-levels structure. Its 2-level structure also causes a larger power consumption with respect to all other solutions, caused by the need to fetch every line of cache 2 times inside the cluster (i.e., both from L1 and L1.5). The baseline two-level cache has the best energy efficiency, 7\% and 4\% better than private cache and single-port shared cache, respectively. It is interesting to note the trade-off between the optimized fetch unit and the legacy one, the one performing better efficiency thanks to the larger 128-bit interface requiring less control overhead for refills, and the other one significantly relaxing the critical path through the core by means of a more straightforward implementation leading to higher operating frequency for the cluster, particularly when scaling up the number of computing cores.
\section{Conclusion}

This work proposes a two-level instruction cache to improve performance while delivering similar energy efficiency with respect to a shared cache, an effective prefetch scheme to reduce performance degradation caused by capacity misses in the small first level of cache, and a timing optimization of the core instruction fetch stage. We explored various instruction cache architectures in an energy-efficient and cost-effective tightly coupled cluster with several signal processing and CNN applications that feature diverse instruction memory access patterns. Results show that the prefetch feature constantly improves the performance up to 7\% while keeping the same energy efficiency in the two-level cache. After timing optimization of the core instruction fetch stage, the two-level cache improves the maximum performance up to 17\% compared with private and shared caches. Finally, the two-level instruction cache with software-enabled prefetch and up to 20\% timing improvement adapts to real-life IoT applications to achieve the highest performance and balanced energy efficiency.


\bibliography{IEEEabrv,bio}

\begin{thebibliography}{10}
\providecommand{\url}[1]{#1}
\csname url@samestyle\endcsname
\providecommand{\newblock}{\relax}
\providecommand{\bibinfo}[2]{#2}
\providecommand{\BIBentrySTDinterwordspacing}{\spaceskip=0pt\relax}
\providecommand{\BIBentryALTinterwordstretchfactor}{4}
\providecommand{\BIBentryALTinterwordspacing}{\spaceskip=\fontdimen2\font plus
\BIBentryALTinterwordstretchfactor\fontdimen3\font minus
  \fontdimen4\font\relax}
\providecommand{\BIBforeignlanguage}[2]{{%
\expandafter\ifx\csname l@#1\endcsname\relax
\typeout{** WARNING: IEEEtran.bst: No hyphenation pattern has been}%
\typeout{** loaded for the language `#1'. Using the pattern for}%
\typeout{** the default language instead.}%
\else
\language=\csname l@#1\endcsname
\fi
#2}}
\providecommand{\BIBdecl}{\relax}
\BIBdecl

\bibitem{8264734}
A.~Loquercio \emph{et~al.}, ``Dronet: Learning to fly by driving,'' \emph{IEEE
  Robotics and Automation Letters}, vol.~3, no.~2, pp. 1088--1095, 2018.

\bibitem{b6}
D.~{Rossi} \emph{et~al.}, ``Energy-efficient near-threshold parallel computing:
  The pulpv2 cluster,'' \emph{IEEE Micro}, vol.~37, no.~5, pp. 20--31, Sep.
  2017.

\bibitem{karam2009trends}
L.~Karam \emph{et~al.}, ``{Trends in multicore DSP platforms},'' \emph{IEEE
  signal processing magazine}, vol.~26, no.~6, pp. 38--49, 2009.

\bibitem{flynn1972some}
M.~J. Flynn, ``{Some computer organizations and their effectiveness},''
  \emph{IEEE transactions on computers}, vol. 100, no.~9, pp. 948--960, 1972.

\bibitem{b20}
D.~Rossi \emph{et~al.}, ``A 60 gops/w, -1.8v to 0.9v body bias ulp cluster in
  28nm utbb fd-soi technology,'' \emph{Solid-State Electronics}, vol. 117, 11
  2015.

\bibitem{b15}
P.~{Meinerzhagen} \emph{et~al.}, ``Benchmarking of standard-cell based memories
  in the sub-$v_{\rm t}$domain in 65-nm cmos technology,'' \emph{IEEE Journal
  on Emerging and Selected Topics in Circuits and Systems}, vol.~1, no.~2, pp.
  173--182, June 2011.

\bibitem{b7}
\BIBentryALTinterwordspacing
I.~Loi \emph{et~al.}, ``Exploring multi-banked shared-l1 program cache on
  ultra-low power, tightly coupled processor clusters,'' in \emph{Proceedings
  of the 12th ACM International Conference on Computing Frontiers}, ser. CF
  '15.\hskip 1em plus 0.5em minus 0.4em\relax New York, NY, USA: ACM, 2015, pp.
  64:1--64:8. [Online]. Available:
  \url{http://doi.acm.org/10.1145/2742854.2747288}
\BIBentrySTDinterwordspacing

\bibitem{b19}
F.~{Oboril} \emph{et~al.}, ``Evaluation of hybrid memory technologies using
  sot-mram for on-chip cache hierarchy,'' \emph{IEEE Transactions on
  Computer-Aided Design of Integrated Circuits and Systems}, vol.~34, no.~3,
  pp. 367--380, March 2015.

\bibitem{8721116}
K.~Kuan and T.~Adegbija, ``Halls: An energy-efficient highly adaptable last
  level stt-ram cache for multicore systems,'' \emph{IEEE Transactions on
  Computers}, vol.~68, no.~11, pp. 1623--1634, 2019.

\bibitem{b21}
J.~{Myers} \emph{et~al.}, ``A subthreshold arm cortex-m0+ subsystem in 65 nm
  cmos for wsn applications with 14 power domains, 10t sram, and integrated
  voltage regulator,'' \emph{IEEE Journal of Solid-State Circuits}, vol.~51,
  no.~1, pp. 31--44, Jan 2016.

\bibitem{b22}
N.~{Ickes} \emph{et~al.}, ``A 10 pj/cycle ultra-low-voltage 32-bit
  microprocessor system-on-chip,'' in \emph{2011 Proceedings of the ESSCIRC
  (ESSCIRC)}, Sep. 2011, pp. 159--162.

\bibitem{b16}
A.~Teman \emph{et~al.}, ``Controlled placement of standard cell memory arrays
  for high density and low power in 28nm fd-soi,'' in \emph{The 20th Asia and
  South Pacific Design Automation Conference}, 2015, pp. 81--86.

\bibitem{origin}
\BIBentryALTinterwordspacing
A.~J. Smith, ``Cache memories,'' \emph{ACM Comput. Surv.}, vol.~14, no.~3, p.
  473–530, Sep. 1982. [Online]. Available:
  \url{https://doi.org/10.1145/356887.356892}
\BIBentrySTDinterwordspacing

\bibitem{Liu_2004}
C.~Liu \emph{et~al.}, ``Organizing the last line of defense before hitting the
  memory wall for cmps,'' in \emph{10th International Symposium on High
  Performance Computer Architecture (HPCA'04)}, 2004, pp. 176--185.

\bibitem{gpu}
H.~Wong \emph{et~al.}, ``Demystifying gpu microarchitecture through
  microbenchmarking,'' in \emph{2010 IEEE International Symposium on
  Performance Analysis of Systems Software}, 2010, pp. 235--246.

\bibitem{Zhang_2005}
\BIBentryALTinterwordspacing
C.~Zhang, F.~Vahid, and W.~Najjar, ``A highly configurable cache for low energy
  embedded systems,'' \emph{ACM Trans. Embed. Comput. Syst.}, vol.~4, no.~2, p.
  363–387, may 2005. [Online]. Available:
  \url{https://doi.org/10.1145/1067915.1067921}
\BIBentrySTDinterwordspacing

\bibitem{Wang_2011}
W.~Wang, P.~Mishra, and S.~Ranka, ``Dynamic cache reconfiguration and
  partitioning for energy optimization in real-time multi-core systems,'' in
  \emph{2011 48th ACM/EDAC/IEEE Design Automation Conference (DAC)}, 2011, pp.
  948--953.

\bibitem{Reinman_1999}
G.~Reinman, B.~Calder, and T.~Austin, ``Fetch directed instruction
  prefetching,'' in \emph{Proceedings of the 32nd Annual ACM/IEEE International
  Symposium on Microarchitecture}, 1999, pp. 16--27.

\bibitem{Ferdman_2008}
M.~Ferdman \emph{et~al.}, ``Temporal instruction fetch streaming,'' in
  \emph{2008 41st IEEE/ACM International Symposium on Microarchitecture}, 2008,
  pp. 1--10.

\bibitem{Ferdman_2011}
M.~Ferdman, C.~Kaynak, and B.~Falsafi, ``Proactive instruction fetch,'' in
  \emph{2011 44th Annual IEEE/ACM International Symposium on Microarchitecture
  (MICRO)}, 2011, pp. 152--162.

\bibitem{Kolli_2013}
A.~Kolli \emph{et~al.}, ``Rdip: Return-address-stack directed instruction
  prefetching,'' in \emph{2013 46th Annual IEEE/ACM International Symposium on
  Microarchitecture (MICRO)}, 2013, pp. 260--271.

\bibitem{Ishii_2021}
Y.~Ishii \emph{et~al.}, ``Re-establishing fetch-directed instruction
  prefetching: An industry perspective,'' in \emph{2021 IEEE International
  Symposium on Performance Analysis of Systems and Software}, pp. 172--182.

\bibitem{diaz2021near}
J.~D{\'\i}az \emph{et~al.}, ``{Near-optimal replacement policies for shared
  caches in multicore processors},'' \emph{The Journal of Supercomputing},
  vol.~77, no.~10, pp. 11\,756--11\,785, 2021.

\bibitem{ghosh2021srcp}
S.~N. Ghosh \emph{et~al.}, ``{SRCP: sharing and reuse-aware replacement policy
  for the partitioned cache in multicore systems},'' \emph{Design Automation
  for Embedded Systems}, vol.~25, no.~3, pp. 193--211, 2021.

\bibitem{xiao2022cache}
J.~Xiao, Y.~Shen, and A.~D. Pimentel, ``{Cache Interference-aware Task
  Partitioning for Non-preemptive Real-time Multi-core Systems},'' \emph{ACM
  Transactions on Embedded Computing Systems (TECS)}, vol.~21, no.~3, pp.
  1--28, 2022.

\bibitem{cabo2021safesu}
G.~Cabo \emph{et~al.}, ``{SafeSU: an extended statistics unit for multicore
  timing interference},'' in \emph{2021 IEEE European Test Symposium
  (ETS)}.\hskip 1em plus 0.5em minus 0.4em\relax IEEE, 2021, pp. 1--4.

\bibitem{b11}
M.~{Gautschi} \emph{et~al.}, ``Near-threshold risc-v core with dsp extensions
  for scalable iot endpoint devices,'' \emph{IEEE Transactions on Very Large
  Scale Integration (VLSI) Systems}, vol.~25, no.~10, pp. 2700--2713, Oct 2017.

\bibitem{b14}
L.~{Benini}, E.~{Flamand}, D.~{Fuin}, and D.~{Melpignano}, ``P2012: Building an
  ecosystem for a scalable, modular and high-efficiency embedded computing
  accelerator,'' in \emph{2012 Design, Automation Test in Europe Conference
  Exhibition (DATE)}, March 2012, pp. 983--987.

\bibitem{Rahimi_2011}
A.~Rahimi \emph{et~al.}, ``A fully-synthesizable single-cycle interconnection
  network for shared-l1 processor clusters,'' in \emph{2011 Design, Automation
  Test in Europe}, 2011, pp. 1--6.

\bibitem{b4}
\BIBentryALTinterwordspacing
D.~Rossi \emph{et~al.}, ``Ultra-low-latency lightweight dma for tightly coupled
  multi-core clusters,'' in \emph{Proceedings of the 11th ACM Conference on
  Computing Frontiers}, ser. CF '14.\hskip 1em plus 0.5em minus 0.4em\relax New
  York, NY, USA: Association for Computing Machinery, 2014. [Online].
  Available: \url{https://doi.org/10.1145/2597917.2597922}
\BIBentrySTDinterwordspacing

\bibitem{b1}
I.~{Loi} \emph{et~al.}, ``The quest for energy-efficient i\$ design in
  ultra-low-power clustered many-cores,'' \emph{IEEE Transactions on
  Multi-Scale Computing Systems}, vol.~4, no.~2, pp. 99--112, April 2018.

\bibitem{DATE_Jie}
C.~Jie \emph{et~al.}, ``Energy-efficient two-level instruction cache design for
  an ultra-low-power multi-core cluster,'' in \emph{Proceedings of the 23rd
  Conference on Design, Automation and Test in Europe}, ser. DATE '20.\hskip
  1em plus 0.5em minus 0.4em\relax San Jose, CA, USA: EDA Consortium, 2020, p.
  1734–1739.

\bibitem{GAP9}
{GreenWaves Technologies}, ``Gap9 product brief,'' 2022,
  \url{https://greenwaves-technologies.com/wp-content/uploads/2022/06/Product-Brief-GAP9-Sensors-General-V1\_14.pdf}.

\bibitem{b17}
A.~{Marongiu} \emph{et~al.}, ``Simplifying many-core-based heterogeneous soc
  programming with offload directives,'' \emph{IEEE Transactions on Industrial
  Informatics}, vol.~11, no.~4, pp. 957--967, Aug 2015.

\bibitem{b18}
{GreenWaves Technologies}, ``Gap8 auto-tiler manual,'' 2018,
  \url{https://greenwaves-technologies.com}.

\end{thebibliography}


\begin{thebibliography}{1}

\bibitem{IEEEhowto:kopka}
H.~Kopka and P.~W. Daly, \emph{A Guide to \LaTeX}, 3rd~ed.\hskip 1em plus
  0.5em minus 0.4em\relax Harlow, England: Addison-Wesley, 1999.

\end{thebibliography}

\begin{IEEEbiography}[{\includegraphics[width=1in,height=1.25in,clip,keepaspectratio]{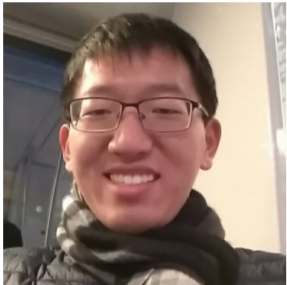}}]{Jie Chen}
received Ph.D. from teh Department of Electrical, Electronic and Information Engineering at the University of Bologna in 2022. At the same, he is working as ASIC designer for Greenwaves Technologies, France. His research activities are currently on ultra-low power multi-core systems, memory systems hierarchy and high speed memory interface.
\end{IEEEbiography}
\vspace{-1cm}

\begin{IEEEbiography}[{\includegraphics[width=1in,height=1.25in,clip,keepaspectratio]{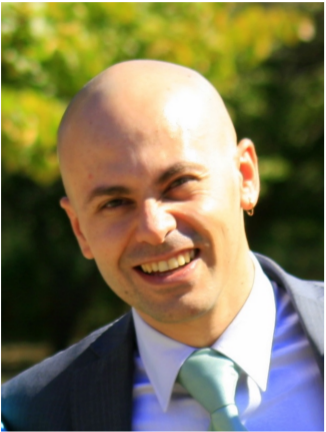}}]{Igor Loi}
received the PhD from the University of Bologna, Italy, in 2010. He has been a post doc researcher in the Department of Electrical, Electronic and Information Engineering at the University of Bologna since 2006. His research activities are currently focused on ultra-low power multi-core systems, memory systems evolution, and ultra low-latency interconnects. In this field he has published more than 40 paper in international peer-reviewed conferences and journals.
\end{IEEEbiography}
\vspace{-1cm}

\begin{IEEEbiography}[{\includegraphics[width=1in,height=1.25in,clip,keepaspectratio]{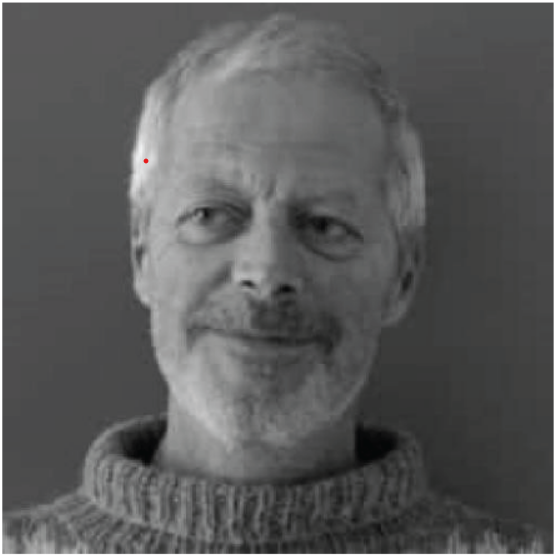}}]{Eric Flamand}
received the Ph.D. degree in computer science from INPG, Grenoble, France, in 1982. He was a Researcher with CNET and CNRS, Grenoble, France on design of low-power parallel processors. He then held different technical management positions within the semiconductor industry with Motorola and ST Microelectronics. He is the co-founder and currently the CTO of Greenwaves Technologies, a French startup developing an IoT processor derived from the PULP project.
\end{IEEEbiography}
\vspace{-1cm}


\begin{IEEEbiography}[{\includegraphics[width=1in,height=1.25in,clip,keepaspectratio]{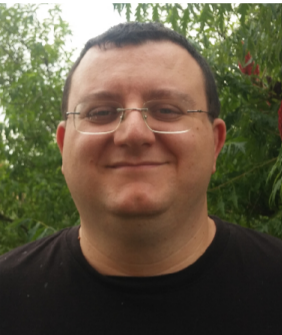}}]{Giuseppe Tagliavini}
received a Ph.D. degree in Electronic Engineering from the University of Bologna, Bologna, Italy, in 2017. He is currently an Assistant Professor with the Department of Computer Science and Engineering (DISI) at the University of Bologna. He has co-authored over 40 papers in international conferences and journals. His main research interests include parallel programming models for embedded systems, run-time and compile-time optimization for multi/many-core platforms, HW/SW co-design of emerging computing.
\end{IEEEbiography}
\vspace{-1cm}

\begin{IEEEbiography}[{\includegraphics[width=1in,height=1.25in,clip,keepaspectratio]{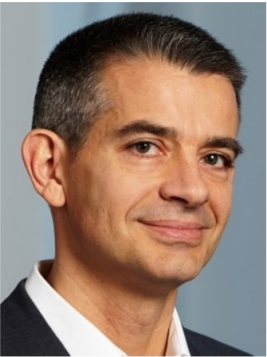}}]{Luca Benini}
received the Ph.D. degree in electrical engineering from Stanford University, Stanford, CA, USA, in 1997. He is currently a Full Professor with the University of Bologna, Bologna, Italy. He has authored over 700 papers in peer-reviewed international journals and conferences, four books, and several book chapters. His current research interests include energy efficient system design and multicore system-on-chip design. Dr. Benini is a member of Academia Europaea. He is currently the Chair of Digital Circuits and Systems with ETH Zürich, Zürich, Switzerland.
\end{IEEEbiography}
\vspace{-1cm}

\begin{IEEEbiography}[{\includegraphics[width=1in,height=1.25in,clip,keepaspectratio]{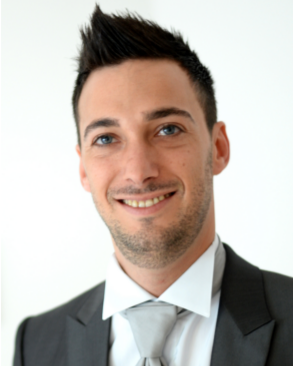}}]{Davide Rossi}
received the Ph.D. degree from the University of Bologna, Bologna, Italy, in 2012. He has been a Post-Doctoral Researcher with the Department of Electrical, Electronic and Information Engineering “Guglielmo Marconi,” University of Bologna, since 2015, where he is currently an Associate Professor. His research interests focus on energy-efficient digital architectures. In this field, he has published more than 100 papers in international peer-reviewed conferences and journals.

\end{IEEEbiography}




\end{document}